%% file: paper_oopsla.tex
\pdfoutput=1
\documentclass[acmsmall, screen]{acmart}

\usepackage{hyperref}
\usepackage{booktabs}
\usepackage{pifont}

\usepackage[title]{appendix}
\usepackage{wrapfig}
\usepackage{enumitem}

\usepackage[normalem]{ulem}

\usepackage{float}

\usepackage{algorithm} 
\usepackage[noend]{algpseudocode} 

\usepackage{amsmath}
\usepackage{stmaryrd}
\usepackage{mathtools}
\usepackage{mathpartir}
\usepackage{xcolor}
\usepackage{csquotes}

\usepackage[capitalize,nameinlink]{cleveref}

\usepackage{tikz}
\usetikzlibrary{automata, positioning, arrows, backgrounds}
\usepackage{pgfplots}
\pgfplotsset{compat=1.18}

\usepackage{euflag_georg}

\definecolor{symhighlight}{HTML}{974006}
\newcommand\syn[1]{\texttt{\color{symhighlight}{#1}}}

\newcommand{\sand}{\mbox{$\, \color{symhighlight}{\pmb{\wedge}} \,$}}

\newcommand{\strue}{\mbox{${\sf \color{symhighlight}{true}}$}}

\newcommand{\simp}{\mbox{$\, \color{symhighlight}{\pmb{\to}} \,$}}

\newcommand{\sall}{\mbox{$\color{symhighlight}{\pmb{\forall}}$}}
\newcommand{\sex}{\mbox{$\color{symhighlight}{\pmb{\exists}}$}}

\newcommand{\sset}[2]{#1 \mathbin{\color{symhighlight}{\pmb{\coloneqq}}} #2}
\newcommand{\srand}[1]{\sset{#1}{{\color{symhighlight}\ast}}}
\newcommand{\snot}{\mbox{$\color{symhighlight}{\pmb{\neg}}$}}

\newcommand\den[1]{\ensuremath{\llbracket#1\rrbracket}}

\setcopyright{rightsretained}
\acmDOI{10.1145/3689761}
\acmYear{2024}
\acmJournal{PACMPL}
\acmVolume{8}
\acmNumber{OOPSLA2}
\acmArticle{321}
\acmMonth{10}
\acmSubmissionID{oopslab24main-p490-p}
\received{2024-04-06}
\received[accepted]{2024-08-18}

\begin{document}

\begin{abstract}
  Many important hyperproperties, such as refinement and generalized non-interference, fall into the class of $\forall\exists$ hyperproperties and require, for each execution trace of a system, the existence of another trace relating to the first one in a certain way.
  The alternation of quantifiers renders $\forall\exists$ hyperproperties extremely difficult to verify, or even just to test. Indeed, contrary to trace properties, where it suffices to find a single counterexample trace, refuting a $\forall\exists$ hyperproperty requires not only to find a trace, but also a proof that no second trace satisfies the specified relation with the first trace.
  As a consequence, automated testing of $\forall\exists$ hyperproperties falls out of the scope of existing automated testing tools.
  In this paper, we present a fully automated approach to detect violations of $\forall\exists$ hyperproperties in software systems.
  Our approach extends bug-finding techniques based on symbolic execution with support for trace quantification.
  We provide a prototype implementation of our approach, and demonstrate its effectiveness on a set of challenging examples.
\end{abstract}

\title{Finding $\forall\exists$ Hyperbugs using Symbolic Execution}

\author{Arthur Correnson}
\orcid{0000-0003-2307-2296}
\affiliation{%
\institution{CISPA Helmholtz Center for Information Security}%
\city{Saarbrücken}%
\country{Germany}}
\email{arthur.correnson@cispa.de}
\authornote{Both authors contributed equally to the paper.}

\author{Tobias Nießen}
\orcid{0000-0002-7712-0006}
\affiliation{%
\institution{TU Wien}%
\city{Vienna}%
\country{Austria}}
\email{tobias.niessen@tuwien.ac.at}
\authornotemark[1]

\author{Bernd Finkbeiner}
\orcid{0000-0002-4280-8441}
\affiliation{%
\institution{CISPA Helmholtz Center for Information Security}%
\city{Saarbrücken}%
\country{Germany}}
\email{finkbeiner@cispa.de}

\author{Georg Weissenbacher}
\orcid{0000-0002-0143-632X}
\affiliation{%
\institution{TU Wien}%
\city{Vienna}%
\country{Austria}}
\email{georg.weissenbacher@tuwien.ac.at}

\keywords{Hyperproperties, Symbolic execution, Bounded model checking, Infinite-state systems}

\ccsdesc[500]{Theory of computation~Logic and verification}
\ccsdesc[300]{Theory of computation~Program specifications}
\ccsdesc[500]{Theory of computation~Program analysis}
\ccsdesc[500]{Software and its engineering~Software defect analysis}
\ccsdesc[500]{Software and its engineering~Formal software verification}

\maketitle

\section{Introduction}

Hyperproperties relate multiple executions of a system. While most initial interest in hyperproperties came from the area of information-flow security~\cite{clarkson_hyperproperties_2010}, where hyperproperties capture important policies like noninference and observational determinism~\cite{clarkson_temporal_2014,mclean_1994}, hyperproperties have also found numerous applications in areas from embedded controllers~\cite{barth_doping_2016} to sorting
algorithms~\cite{chaudhuri_robustness_2012}.

A key benefit of hyperproperties is that they can capture general concepts, such as \emph{symmetry}, that are expected to hold in wide variety of situations.
Consider, as an example, the two versions of the voting protocol shown in \cref{fig:voting}.

\begin{figure}[ht!]
  \centering
  \begin{minipage}{.4\textwidth}
    \begin{tabular}{l}
      $\mathit{countA} \gets 0$\\
      $\mathit{countB} \gets 0$\\
      \textbf{loop} \\
      \quad \textbf{input} $\mathit{vote} \in \{A, B\}$\\
      \quad \textbf{if} $\mathit{vote} = A$ \textbf{then}\\
      \quad\quad $\mathit{countA} \gets \mathit{countA} + 1$\\
      \quad \textbf{else}\\
      \quad\quad $\mathit{countB} \gets \mathit{countB} + 1$\\
      \quad \textbf{output} $\mathit{countA}, \mathit{countB}$
    \end{tabular}
  \end{minipage}%
  \begin{minipage}{.4\textwidth}
    \begin{tabular}{l}
      $\mathit{countA} \gets 0$\\
      $\mathit{countB} \gets 0$\\
      \textbf{loop} \\
      \quad \textbf{input} $\mathit{vote} \in \{A, B\}$\\
      \quad \textbf{if} $\mathit{vote} = A$ \textbf{then}\\
      \quad\quad $\mathit{countA} \gets \mathit{countA} + 1$\\
      \quad \textbf{else}\\
      \quad\quad $\mathit{countB} \gets {\underline{\mathit{countA}}} + 1$\\
      \quad \textbf{output} $\mathit{countA}, \mathit{countB}$
    \end{tabular}
  \end{minipage}
  \caption{Two versions of a simple voting protocol with a bug underlined}
  \label{fig:voting}
\end{figure}

The version on the left correctly tallies the votes of the two candidates $A$ and $B$. The version on the right contains a bug, where $\mathit{countB}$ is set to $\mathit{countA} + 1$ instead of the correct $\mathit{countB} + 1$. The fact that the version on the right cannot be correct can easily be seen, without even specifying the functionality of the protocol, by noticing that the buggy version treats $A$ differently from $B$.

More formally, the problem with the version of the protocol on the right is revealed by checking the hyperproperty specified by the following temporal formula: 
$$\forall \pi_1. \exists \pi_2. \square (\mathit{countA}_{\pi_1} = \mathit{countB}_{\pi_2} \wedge \mathit{countA}_{\pi_2} = \mathit{countB}_{\pi_1})$$

The formula specifies that for every execution $\pi_1$, there must exist an execution $\pi_2$ such that the counts for $A$ and $B$ are exactly flipped compared to $\pi_1$.
In other words, through the election process, there should be an opportunity for $A$ to receive exactly the same votes as $B$ (and vice versa).
Clearly, the voting protocol on the right violates this property, since any vote for $B$ has the effect of making $B$ beat $A$ by one vote.

Much of the research on the verification of hyperproperties has focused on \emph{proving} that a hyperproperty is satisfied. Techniques for showing \emph{violations} of hyperproperties, on the other hand, are either restricted to the analysis of finite-state systems \cite{groote_bounded_2021}, limited to the testing of a fixed property \cite{lesly_spectre_2021,lesly_binsec_2023}, or require human guidance to find errors~\cite{dickerson_rhle_2022}. This is unfortunate, since the detection of bugs and design errors is of great help during software development~\cite{godefroid_bugs_2005}.
In examples like the voting protocol, we are not so much interested in proving that the version on the left satisfies symmetry (which, by itself, does not guarantee that the protocol is functionally correct) than in finding the violation of symmetry in the version on the right, which immediately establishes that the protocol \emph{cannot} be correct.

In this paper, we develop new foundations for fully automated \enquote{hyperbug} finding (i.e., the detection of hyperproperty violations) in software systems. Contrary to existing approaches, ours does not require human intervention. Further, it is capable of providing concrete counterexamples demonstrating why a hyperproperty does not hold.

\paragraph{Challenges.}

Hyperproperties are generally classified based on the type of trace quantification that they require. For example, $k$-safety properties are properties that universally quantify on $k$ concurrent traces of a system and express a relation between them.
Verifying (or testing) $k$-safety properties can be reduced to the analyzing single trace properties on bigger systems obtained by self-composition \cite{barthe_secure_2004}.
It is then possible to exploit symmetries in the resulting composed system (or in the property itself) to drastically speed up the verification and bug-finding algorithms \cite{farzan_automated_2019, farzan_reduction_2019, lesly_binsec_2023}.
Other properties, often referred to as \enquote{hyperliveness} properties, can only be expressed with an alternation of universal and existential quantifiers.
Such properties require for every trace the existence of another trace that relates to the first one in a certain way. Important examples of such $\forall\exists$ hyperproperties include refinement, generalized non-interference~\cite{mccullough_noninterference_1988}, and delimited information release~\cite{sabelfeld_model_2004}.

Verifying $\forall\exists$ hyperproperties cannot be reduced to the verification of simpler trace properties by self-composition. For every universally quantified trace, a corresponding existential witness has to be searched for.
Detecting violations of $\forall\exists$ properties is also complex: it requires to find both a trace and a proof that no second trace is compatible.
Again, this generally requires to enumerate all possible combinations of traces.
In the context of software systems, the number of traces is generally infinite.
Popular bug-finding methods such as symbolic execution~\cite{cadar_klee_2008,godefroid_dart_2005,sen_cute_2005} and fuzzing~\cite{manes_art_2021} overcome this challenge by exploring only a subset of all possible traces of a system.
Such under-approximating methods cannot directly be applied to $\forall\exists$ hyperproperties as considering only a strict subset of all possible traces is not sufficient to disprove the existence of a witness trace for the inner-most existential quantifier.

\paragraph{Contributions.}

In this paper, we present the first symbolic execution method that checks automatically whether one or more programs satisfy a given $\forall\exists$ hyperproperty. The key idea of our approach is to combine two symbolic execution engines: one to find a universal trace, and one to encode the fact that no matching trace exists. 

We consider properties expressed in a fragment of OHyperLTL~\cite{beutner_software_2022}, a temporal logic well-suited for specifying hyperproperties of reactive systems.
Importantly, the fragment of $\textrm{OHyperLTL}$ we choose, which we refer to as $\textrm{OHyperLTL}_\mathit{safe}$, can express a wide range of $\forall\exists$ properties.

A key methodological point is to equip $\textrm{OHyperLTL}_\mathit{safe}$ with a bounded semantics, allowing to determine whether a formula is violated by only looking at finite execution prefixes.
This semantics also extends the original semantics of $\textrm{OHyperLTL}$ by making it more suitable for reasoning about systems whose executions may or may not terminate.
Importantly, we prove that the bounded semantics \textit{agrees} with the unbounded one, and we demonstrate that it can express properties of a wider range of systems.

Based on the newly introduced bounded semantics of $\textrm{OHyperLTL}_\mathit{safe}$, we develop an algorithm to automatically detect hyperproperty violations.
This new algorithm extends existing bug-finding approaches based on symbolic execution to support arbitrary trace quantification, and we further optimize this algorithm for specifications that require quantifier alternation as it occurs in $\forall\exists$ hyperproperties.
We also provide a prototype implementation of the proposed algorithm.
We evaluate this prototype on a selection of benchmarks drawn from the relevant literature, as well as new benchmarks specifically designed to test its limitations. Experimental results demonstrate the effectiveness of our algorithm in locating $\forall\exists$ hyperproperty violations, without any human intervention, and within seconds.

\paragraph{Note.}
Throughout this paper, we use \textcolor{symhighlight}{this color} to mark symbolic tokens, i.e., syntactic elements that are consumed or produced by any mathematical formulations in this paper.

\section{Preliminaries}
\label{sec:prelim}

\paragraph{First-order logic and theories~\cite{barwise_introduction_1977}.}
We fix some arbitrary underlying first-order theory $\mathcal{T}$ with domain $\operatorname{Val}_\mathcal{T}$, and we use the following notations and conventions:

\begin{itemize}
  \item $\operatorname{Term}_\mathcal{T}(X)$ is the set of first-order terms over the set of variables $X$.
  \item $\operatorname{Form}_\mathcal{T}(X)$ is the set of first-order formulas over the set of variables $X$.
  \item For any term $t \in \operatorname{Term}_\mathcal{T}(X)$, and any variable assignment $\rho \in (\operatorname{Val}_\mathcal{T})^X$, we denote by $\den{t}^\rho_\mathcal{T} \in \operatorname{Val}_\mathcal{T}$ the value of $t$.
  \item For any formula $\varphi \in \operatorname{Form}_\mathcal{T}(X)$, and any variable assignment $\rho \in (\operatorname{Val}_\mathcal{T})^X$, we write $\rho \models_\mathcal{T} \varphi$ if and only if $\rho$ is a model of $\varphi$ in $\mathcal{T}$.
  \item For (not necessarily disjoint) sets of variables $X$ and $X'$, $\sigma \in \operatorname{Term}(X')^X$ is a substitution.
  \item For $t \in \operatorname{Term}_\mathcal{T}(X)$ and $\sigma \in \operatorname{Term}(X')^X$ $\varphi[/\sigma] \in \operatorname{Term}_\mathcal{T}(X)$ is the term obtained from $t$ by substituting every free variables $x$ with the term $\sigma(x)$.
  \item For $\varphi \in \operatorname{Form}_\mathcal{T}(X)$ and $\sigma \in \operatorname{Term}(X')^X$ $\varphi[/\sigma] \in \operatorname{Form}_\mathcal{T}(X)$ is the formula obtained from $\varphi$ by substituting every free variables $x$ with the term $\sigma(x)$.
\end{itemize}

\paragraph{Program Graphs.} Throughout this paper, we model programs as program graphs to facilitate reasoning and simplify formal definitions. Program graphs operate on a finite set $X$ of program variables and have a finite set of vertices called program locations.
Edges of a program graph are labeled with \textit{guarded variable assignments} representing conditional updates of the program variables. Assignments are either of the form $\sset{x}{e}$ where $x \in X$ and $e \in \operatorname{Term}_\mathcal{T}(X)$ (assignment of an expression to a variable) or of the form $\srand{x}$ (nondeterministic assignments/user inputs). We note $\mathit{Instr}$ the set of possible assignment instructions (i.e., $\mathit{Instr} = \{ \sset{x}{e} \mid x \in X, e \in \operatorname{Term}_\mathcal{T}(X) \} \cup \{ \srand{x}\mid x \in X \}$). Formally, program graphs are defined as follows:

\begin{definition}[Program graph]\label{def:program-graph}
  A \emph{program graph} (or \emph{control flow graph}) is a tuple \[
    G = (\langle\mathcal{L}, \mathcal{E}\rangle, \ell_0, \mathit{effect}, \mathit{guard})
  \] where \begin{itemize}
  \item $\mathcal{L}$ is a non-empty, finite set of program locations,
  \item $\mathcal{E} \subseteq \mathcal{L} \times \mathcal{L}$ is a set of edges connecting locations,
  \item $\ell_0 \in \mathcal{L}$ is the initial program location,
  \item $\mathit{effect} : \mathcal{E} \to \mathit{Instr}$ maps each edge to an instruction, and
  \item $\mathit{guard} : \mathcal{E} \to \operatorname{Form}_\mathcal{T}(X)$ maps edges to
  (quantifier-free) formulas over program variables $X$.
\end{itemize}
\end{definition}

For readability, we represent program graphs using diagrams. \cref{fig:example-graph} shows how a simple program can be represented as a diagram.

\tikzset{->,shorten >=1pt,auto,node distance=2.5cm,on grid,initial text=,
    every state/.style={minimum size=20pt,inner sep=0pt},
    every node/.style={font=\normalsize}}

\begin{figure}[ht!]
  \begin{center}
    \begin{minipage}{.4\textwidth}
      \begin{tabular}{l}
        \textbf{loop}\\
        \quad \textbf{input} $x_\textrm{in}$\\
        \quad \textbf{if} $x_\textrm{in} > 0$ \textbf{then}\\
        \quad \quad \textbf{output} $1$\\
        \quad \textbf{else}\\
        \quad \quad \textbf{output} $0$\\
      \end{tabular}
    \end{minipage}
    \begin{minipage}{.4\textwidth}
      \begin{tikzpicture}
        \node[initial, state] (A) {$\ell_0$};
        \node[state, right of = A] (B) {$\ell_1$};
        \draw[->] (A) edge node {$\srand{x_\mathit{in}}$} (B);
        \draw[->, bend left = 70] (B) edge node[below] {$x > 0 \ \syn{:} \ \sset{\mathit{output}}{1}$} (A);
        \draw[->, bend right = 70] (B) edge node[above] {$x \le 0 \ \syn{:} \ \sset{\mathit{output}}{0}$} (A);
      \end{tikzpicture}
    \end{minipage}
  \end{center}
  \caption{A simple program and a possible translation as a program graph}
  \label{fig:example-graph}
  \Description{A simple program and a possible translation as a program graph}
\end{figure}
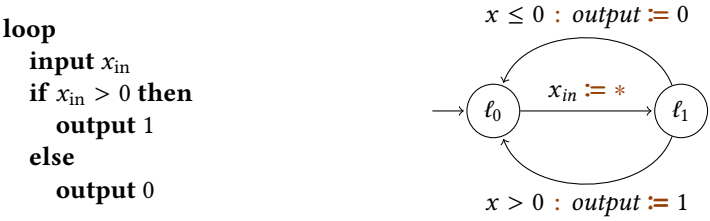

Programs operate on memories $m \in \mathcal{M}$, where $\mathcal{M} \coloneqq (\operatorname{Val}_\mathcal{T})^X$ is the set of all possible assignments of values to program variables.
For simplicity, throughout this paper, we assume that there is a particular initial memory $m_0 \in \mathcal{M}$. For example, if $\operatorname{Val}_\mathcal{T} = \mathbb{N}$, one may initialize the values of all variables to zero.
Given a program graph with locations $\mathcal{L}$, the set of its execution states is $\mathcal{S} = \mathcal{L} \times \mathcal{M}$.
For a state $s = \langle \ell, m \rangle \in \mathcal{S}$, we define $\operatorname{mem}(s) \coloneqq m$.
For an assignment instruction $i$ and memories $m, m' \in \mathcal{M}$, we note $\langle m, i\rangle \Downarrow m'$ if executing the assignment $i$ updates $m$ to $m'$.
\begin{definition}[Semantics of instructions]\label{def:concrete-instruction-semantics}
  \begin{mathpar}
    \inferrule[assign]{ ~ }{\langle m, \sset{x}{e} \rangle \Downarrow m[x \gets \den{e}^m_\mathcal{T}]}\qquad
    \inferrule[havoc]{ v \in \operatorname{Val}_\mathcal{T} }{\langle m, \srand{x}\rangle \Downarrow m[x \gets v]}
  \end{mathpar}
\end{definition}

For a program graph $G$ and two states $s_1, s_2 \in \mathcal{S}$, we note $G \vdash s_1 \hookrightarrow s_2$ if and only if there is a possible transition from state $s_1$ to $s_2$ in $G$.

\begin{definition}[Transition semantics]\label{def:concrete-operational-semantics}
  Let $G = (\langle\mathcal{L}, \mathcal{E}\rangle, \ell_0, \mathit{effect}, \mathit{guard})$ be a program graph. The transition relation $\hookrightarrow$ is defined as follows:
  \begin{mathpar}
    \inferrule[step]{ \langle \ell_1, \ell_2 \rangle \in \mathcal{E}\\m_1 \models_\mathcal{T} \mathit{guard}(\ell_1, \ell_2)\\\langle m_1, \mathit{effect}(\ell_1, \ell_2)\rangle \Downarrow m_2 }{G \vdash \langle \ell_1, m_1 \rangle \hookrightarrow \langle \ell_2, m_2 \rangle}
  \end{mathpar}
\end{definition}

Execution traces of a program $G$ are defined as sequences of valid computation steps starting from the state $\langle \ell_0, m_0 \rangle$, where $\ell_0$ is the initial location of $G$ and $m_0$ is the fixed initial memory.
We use $\mathcal{S}^{*}$, $\mathcal{S}^{\omega}$, and $\mathcal{S}^{\infty} = \mathcal{S}^{*} \cup \mathcal{S}^{\omega}$ to denote the sets of finite, infinite, and mixed traces, respectively.

\begin{definition}[Trace semantics]\label{def:concrete-traces}
  Let $s_0 = \langle \ell_0, m_0 \rangle$ be the initial state and, for any trace $\tau \in \mathcal{S}^\infty$, let $\tau_i$ be the $i$-th pair of location and memory within $\tau$. We define
  \begin{align*}
    \operatorname{Traces}^*(G) &:= \{ \tau \in \mathcal{S}^* \mid \tau_0 = s_0 \wedge G \vdash \tau_i \hookrightarrow \tau_{i + 1}, \forall i \in [0, |\tau| - 1] \}\\
    \operatorname{Traces}^k(G) &:= \{ \tau \in \operatorname{Traces}^*(G) \mid |\tau| = k \}\\
    \operatorname{Traces}^\omega(G) &:= \{ \tau \in \mathcal{S}^\omega \mid \tau_0 = s_0 \wedge G \vdash \tau_i \hookrightarrow \tau_{i + 1}, \forall i \in \mathbb{N} \}\\
  \end{align*}
\end{definition}

\section{Specifying Hyperproperties of Programs in $\textrm{OHyperLTL}_\mathit{safe}$}
\label{sec:ohyperltlsafe}

This paper is concerned with automated testing of hyperproperties. We consider hyperproperties expressed in OHyperLTL~\cite{beutner_software_2022}, a temporal logic for specifying hyperproperties of software systems. OHyperLTL allows universal and existential quantification over traces of a system, and uses temporal operators to express relations between the quantified traces. In this paper, we consider a fragment of OHyperLTL in which the relations are restricted to (relational) invariants. We call this fragment $\textrm{OHyperLTL}_\mathit{safe}$. This section introduces the syntax and the semantics of $\textrm{OHyperLTL}_\mathit{safe}$.

\subsection{Syntax}

An $\textrm{OHyperLTL}_\mathit{safe}$ formula begins with a sequence of universal and existential trace quantifiers $Q^{G_i}\pi_i : O_i$, where $Q \in \{\, \forall, \exists \,\}$, each quantifying over the traces $\pi_i$ of a program $G_i$. Additionally, each quantifier specifies a set of locations $O_i$ at which traces should be observed.
Such observation points serve the purpose of synchronizing the different traces and deciding when they should be compared.
The remainder of the formula specifies an invariant $\square\varphi$ that should always hold across all quantified traces.

\begin{definition}[Syntax of $\textrm{OHyperLTL}_\mathit{safe}$]
  The syntax of $\textrm{OHyperLTL}_\mathit{safe}$ is defined as follows:
  \[
    \begin{array}{ll}
      \psi &::= \forall^G \pi : O.\psi \mid \exists^G \pi : O.\psi \mid \square \varphi,
    \end{array}
  \]
  where $\pi$ is a trace variable drawn from a set $\mathcal{V}$, and $\varphi \in \operatorname{Form}_\mathcal{T}(X_\mathcal{V})$ is a quantifier-free first-order formula over $\mathcal{T}$ with free variables in $X_\mathcal{V} = \bigcup_{\pi \in \mathcal{V}} X_\pi$, i.e., for any program variable $x \in X$ and any trace variable $\pi \in \mathcal{V}$, $x_\pi \in X_\mathcal{V}$.
\end{definition}

\begin{figure}[h!]
  \begin{minipage}{.4\textwidth}
    Program $G$:
    \begin{center}
      \begin{tabular}{l}
        \textbf{loop} \\
        \quad \textbf{inputs} $\mathit{pub}, \mathit{sec} \in \mathbb{Z}$\\
        \quad \textbf{havoc} $r \in \mathbb{Z}$\\
        \quad \textbf{output} $\mathit{sec} + r$
      \end{tabular}
    \end{center}
  \end{minipage}
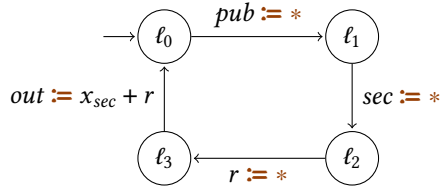
  \begin{minipage}{.5\textwidth}
    \begin{center}
      \begin{tikzpicture}
        \node[initial, state] (A) {$\ell_0$};
        \node[state, right of = A] (B) {$\ell_1$};
        \node[state, below of = B, yshift=.9cm] (C) {$\ell_2$};
        \node[state, below of = A, yshift=.9cm] (D) {$\ell_3$};
        \draw[->] (A) edge node {$\srand{\mathit{pub}}$} (B);
        \draw[->] (B) edge node {$\srand{\mathit{sec}}$} (C);
        \draw[->] (C) edge node {$\srand{r}$} (D);
        \draw[->] (D) edge node {$\sset{\mathit{out}}{x_\mathit{sec} + r}$} (A);
      \end{tikzpicture}
    \end{center}
  \end{minipage}
  \bigskip
  \begin{center}
    Specification of GNI
    \fbox{$
      \forall^G \pi_1 : \{ \ell_0 \}. \forall^G \pi_2 : \{ \ell_0 \}. \exists^G \pi_3 : \{ \ell_0 \}. \square (\mathit{pub}_{\pi_1} = \mathit{pub}_{\pi_3} \wedge \mathit{out}_{\pi_1} = \mathit{out}_{\pi_3} \wedge \mathit{sec}_{\pi_2} = \mathit{sec}_{\pi_3})
    $}
  \end{center}
  \caption{A simple program, and a specification of GNI in OHyperLTL}
  \label{fig:example-gni}
\end{figure}

\cref{fig:example-gni} shows how generalized non-interference, an important security property, can be precisely expressed in $\textrm{OHyperLTL}_\mathit{safe}$ for a simple program.
Generalized non-interference (GNI) ensures that even when a difference in secret inputs causes the outputs of the system to change (potentially revealing secret data), we could have observed the same output difference with any another secret inputs~\cite{mccullough_noninterference_1988}.
Thus, difference of outputs does not provide any information on secret inputs.

We note that the ability to choose observation points is a necessary feature to synchronize multiple traces evolving at different paces~\cite{beutner_software_2022}.
For example, the following two programs both compute the double of their input, but the program on the right takes more computation steps to achieve the goal. Observation points allow to synchronize traces of both programs when they reach the $\textbf{output}$ instruction, and to ignore intermediate computation steps.

\begin{center}
  \vspace{1em}
  \begin{minipage}{.4\textwidth}
    \begin{tabular}{l}
      \textbf{loop}\\
      \quad \textbf{inputs} $x \in \mathbb{Z}$\\
      \quad $y \gets 2x$\\
      \quad \textbf{output} $y$\\
      \quad
    \end{tabular}
  \end{minipage}%
  \begin{minipage}{.4\textwidth}
    \begin{tabular}{l}
      \textbf{loop}\\
      \quad \textbf{inputs} $x \in \mathbb{Z}$\\
      \quad $y \gets x$\\
      \quad $y \gets y + x$\\
      \quad \textbf{output} $y$
    \end{tabular}
  \end{minipage}
  \vspace{1em}
\end{center}

Another important feature of OHyperLTL is that trace quantification is relative to user-specified programs. This allows to draw execution traces from different programs to express relational hyperproperties such as refinement.
For example, \cref{fig:example-refinement} shows how to specify that a program \textsc{min} calculating the minimum of two integers $x$ and $y$ is a \textit{refinement} of a program \textsc{flip} that nondeterministically selects $x$ or $y$.

\begin{figure}[h!]
  \begin{minipage}{.4\textwidth}
    Program \textsc{min}:
    \begin{center}
      \begin{tabular}{l}
        \textbf{loop}\\
        \quad \textbf{inputs} $x, y \in \mathbb{Z}$\\
        \quad \textbf{if} $x < y$ \textbf{then}\\
        \quad \quad \textbf{output} $x$\\
        \quad \textbf{else}\\
        \quad \quad \textbf{output} $y$\\
      \end{tabular}
    \end{center}
  \end{minipage}
  \begin{minipage}{.4\textwidth}
    \begin{center}
      \begin{tikzpicture}
        \node[initial, state] (A) {$\ell_0$};
        \node[state, right of = A, xshift=-.8cm] (B) {$\ell_1$};
        \node[state, right of = B, xshift=-.8cm] (C) {$\ell_2$};
        \draw[->] (A) edge node {$\srand{x}$} (B);
        \draw[->] (B) edge node {$\srand{y}$} (C);
        \draw[->, bend left = -45, above] (C) edge node {$x < y \ \syn{:} \ \sset{\mathit{out}}{x}$} (A);
        \draw[->, bend left = 45, below] (C) edge node {$x \ge y \ \syn{:} \ \sset{\mathit{out}}{y}$} (A);
      \end{tikzpicture}
    \end{center}
  \end{minipage}
  \begin{minipage}{.4\textwidth}
    \bigskip
    Program \textsc{flip}:
    \begin{center}
      \begin{tabular}{l}
        \textbf{loop}\\
        \quad \textbf{inputs} $x, y \in \mathbb{Z}$\\
        \quad \textbf{either}\\
        \quad \quad \textbf{output} $x$\\
        \quad \textbf{or}\\
        \quad \quad \textbf{output} $y$\\
      \end{tabular}
    \end{center}
  \end{minipage}
  \begin{minipage}{.4\textwidth}
    \bigskip
    \begin{center}
      \begin{tikzpicture}
        \node[initial, state] (A) {$\ell_0$};
        \node[state, right of = A, xshift=-.8cm] (B) {$\ell_1$};
        \node[state, right of = B, xshift=-.8cm] (C) {$\ell_2$};
        \draw[->] (A) edge node {$\srand{x}$} (B);
        \draw[->] (B) edge node {$\srand{y}$} (C);
        \draw[->, bend left = -45, above] (C) edge node {$\sset{\mathit{out}}{x}$} (A);
        \draw[->, bend left = 45, below] (C) edge node {$\sset{\mathit{out}}{y}$} (A);
      \end{tikzpicture}
    \end{center}
  \end{minipage}
  \bigskip
  \begin{center}
    Specification of refinement\\
    \fbox{$
      \forall^\textsc{min} \pi_1 : \{ \ell_0 \}. \exists^\textsc{flip} \pi_2 : \{ \ell_0 \}. \square (x_{\pi_1} = x_{\pi_2} \wedge y_{\pi_1} = y_{\pi_2} \wedge \mathit{out}_{\pi_1} = \mathit{out}_{\pi_2})
    $}
  \end{center}
  \caption{Two programs, and a specification of refinement in OHyperLTL}
  \label{fig:example-refinement}
\end{figure}

As will be apparent once we establish semantics in the following section, the specification in \cref{fig:example-refinement} holds: for every execution of $\textsc{min}$, there exists an execution of $\textsc{flip}$ such that their inputs and outputs match.
However, if the two programs were to be swapped, i.e., if the specification began with the quantifiers $\forall^\textsc{flip} \pi_1 : \{ \ell_0 \}. \exists^\textsc{min} \pi_2 : \{ \ell_0 \}$ instead, the property would clearly not hold.
In other words, $\textsc{flip}$ is not a refinement of $\textsc{min}$.

\subsection{Semantics}

$\textrm{OHyperLTL}_\mathit{safe}$ formulas are evaluated on projections of traces that effectively hide computation steps occurring between two observation points.
Given a set $O$ of \emph{observed locations}, we define the following set of \emph{observed traces}:

\begin{definition}[Observational trace semantics]\label{def:concrete-o-traces}
  \begin{align*}
    \operatorname{Traces}^\omega_O(G) &:= \{ [\sigma]_O \mid \sigma \in \operatorname{Traces}^\omega(G) \wedge |\sigma|_O = \infty \}\\
    \operatorname{Traces}^*_O(G) &:= \{ [\sigma]_O \mid \sigma \in \mathit{Traces}^*(G) \}\\
    \operatorname{Traces}^k_O(G) &:= \{ \sigma \in \operatorname{Traces}^*_O(G) \mid |\sigma| = k \},
  \end{align*}
  where $[\sigma]_O$ is the sequence of states obtained from a trace $\sigma$ by removing all states whose locations are not in $O$, and $|\sigma|_O = |\,[\sigma]_O\,|$ is the number of states in $\sigma$ with a location in $O$.
\end{definition}

\noindent Given a formula $\psi$ and a partial map $\Pi$ mapping the free variables of $\psi$ to concrete (projections of) traces, we write $\Pi \models \psi$ if the traces in $\Pi$ satisfy $\psi$. The semantics is defined as follows:

\begin{definition}[Infinite-trace semantics of $\textrm{HyperLTL}_\mathit{safe}$]\label{def:ohyperltlsafe-inf-semantics}
  \begin{align*}
    \Pi &\models^\omega \forall^G\pi : O.\psi & \iff & \quad \forall \sigma \in \operatorname{Traces}^\omega_O(G), \ \Pi[\pi \gets \sigma] \vDash \psi\\
    \Pi &\models^\omega \exists^G\pi : O.\psi & \iff & \quad  \exists \sigma \in \operatorname{Traces}^\omega_O(G), \ \Pi[\pi \gets \sigma] \vDash \psi\\
    \Pi &\models^\omega \square\varphi & \iff & \quad \forall i \in \mathbb{N}, \ \Pi_i \models_\mathcal{T} \varphi,\\
    &&&\quad\textrm{where} \ \Pi_i(x_\pi) := \operatorname{mem}(\Pi(\pi)_i)(x)\\
    &\models^\omega \psi & \iff & \quad \emptyset \models \psi
  \end{align*}
\end{definition}
\bigskip

It is important to note that this semantics, which closely follows the original definition of OHyperLTL~\cite{beutner_software_2022}, completely disregards finite traces, as well as infinite traces with only finitely many observation points.
This can become a major obstacle when specifying properties of software systems that have both terminating and non-terminating behaviors.
In particular, any OHyperLTL formula starting with $\forall^G$ is trivially satisfied if $G$ is a terminating program or a program whose traces have only finitely many observation points. Dually, any OHyperLTL formula starting with $\exists^G$ is trivially violated for such a program $G$. In other words, under this semantics, universal trace quantification is too weak, and existential trace quantification is too strong.

For example, the following echo server trivially satisfies non-interference under the infinite-trace semantics, even though it clearly leaks the secret input:

\begin{center}
  \vspace{1em}
  \begin{tabular}{l}
    \textbf{repeat} $10$ \textbf{times} \\
    \quad \textbf{inputs} $\mathit{pub}, \mathit{sec} \in \mathbb{Z}$\\
    \quad \textbf{output} $\mathit{sec}$
  \end{tabular}
  \vspace{1em}
\end{center}

Indeed, since the main loop of the server is executed only 10 times, it produces only finite traces. In the following, we progressively refine the semantics of $\textrm{OHyperLTL}_\mathit{safe}$ to address this issue. The goal is to obtain a semantics that coincides with \cref{def:ohyperltlsafe-inf-semantics} for programs with only traces with infinitely many observations, while providing a more intuitive treatment of traces with finitely many observations.

\subsection{Bounded Semantics}

As a first step, we begin by introducing a bounded variant of the semantics that only considers traces with exactly $k$ observations.

\begin{definition}[Bounded semantics of $\textrm{OHyperLTL}_\mathit{safe}$]\label{def:ohyperltlsafe-bounded-semantics}
  \begin{align*}
    \Pi &\models^k \forall^G\pi : O.\psi & \iff & \quad \forall \sigma \in \operatorname{Traces}^O_k(G), \ \Pi[\pi \gets \sigma] \models^k \psi\\
    \Pi &\models^k \exists^G\pi : O.\psi & \iff & \quad  \exists \sigma \in \operatorname{Traces}^O_k(G), \ \Pi[\pi \gets \sigma] \models^k \psi\\
    \Pi &\models^k \square\varphi & \iff & \quad \forall i \in [0, k - 1], \ \Pi_i \models_\mathcal{T} \varphi,\\
    &&&\quad\textrm{where} \ \Pi_i(x_\pi) := \operatorname{mem}(\Pi(\pi)(i))(x)\\
    &\models^k \psi & \iff & \quad \emptyset \models^k \psi
  \end{align*}
\end{definition}

Intuitively, $\models^k \psi$ means that $\psi$ is not violated if we consider only
prefixes of observational length exactly $k$. However, it does not provide any information on traces with strictly less or strictly more observations (more in the next paragraph).
Nonetheless, it can be proved that the bounded semantics agrees with the unbounded one for non-terminating programs that only have traces with infinitely many observations. We call such program \textit{infinitely observable}.

\begin{definition}[Infinitely observable programs]
  Let $G$ be a program graph and $O$ a set of observed locations.
  We say that $G$ is \textit{infinitely observable with respect to $O$} if every trace $\tau \in \operatorname{Traces}^k_O(G)$ can be extended into a trace $\bar{\tau} \in \operatorname{Traces}^\omega_O(G)$. In other words, infinitely observable programs are non-terminating programs such that all their traces have infinitely many observations.
\end{definition}

\noindent
This definition is naturally extended to $\textrm{HyperLTL}_\mathit{safe}$ specifications as follows:

\begin{definition}[Infinitely observable specifications]
  Let $\psi$ be a $\textrm{HyperLTL}_\mathit{safe}$ formula. We say that $\psi$ is \textit{infinitely observable} if, for all quantified program $G$ annotated with observed locations $O$ in $\psi$, $G$ is infinitely observable with respect to $O$.
\end{definition}

\begin{theorem}
  Let $\psi$ be an infinitely observable specification and $k \in \mathbb{N}$. Then $\models^\omega \psi \implies \models^k \psi$.
  \label{thm:agree}
\end{theorem}

\newenvironment{proofsketch}
  {\renewcommand*{\proofname}{Proof sketch}\begin{proof}}
  {\end{proof}\renewcommand*{\proofname}{Proof}}

\begin{proofsketch}
  The proof goes by induction on the formula $\psi$ for arbitrary mappings $\Pi \in (\mathcal{S}^\omega)^\mathcal{V}$ and $\Pi' \in (\mathcal{S}^k)^\mathcal{V}$ such that $\Pi'(\pi)$ is a prefix of $\Pi(\pi)$ for every $\pi \in \mathcal{V}$ (we note $\Pi' \sqsubset \Pi$). The case where $\psi$ is of the form $\square \varphi$ is straightforward. The difficult cases are when $\psi$ is of the form $\forall^G \pi : O, \psi'$ or $\exists^G \pi : O, \psi'$: \begin{itemize}
    \item Suppose $\psi = \forall^G \pi : O. \psi'$, and $\Pi \models^\omega \psi$.
    We have to show $\Pi' \models^k \psi$. Let $\tau \in \operatorname{Traces}^k_O(G)$, it is enough to show $\Pi'[\pi \gets \tau] \models^k \psi'$.
    Since $G$ is infinitely observable $\tau$ can be extended to $\bar{\tau} \in \operatorname{Traces}^\omega(G)$, and since $\Pi \models^\omega \psi$, $\Pi[\pi \gets \bar{\tau}] \models^\omega \psi'$. Further, since $\Pi' \sqsubset \Pi$, it is easy to see that $\Pi'[\pi \gets \tau] \sqsubset \Pi[\pi \gets \bar{\tau}]$. By induction hypothesis, it follows that $\Pi[\pi \gets \tau] \models^k \psi'$.
    \item Suppose $\psi = \exists^G \pi : O. \psi'$, and $\Pi \models^\omega \psi$.
    We have to show $\Pi' \models^k \psi'$. By definition of the bounded semantics, it is enough to find some $\tau \in \operatorname{Traces}^k_O(G)$ such that $\Pi'[\pi \gets \tau] \models^k \psi'$.
    Since $\Pi \models^\omega \psi'$, there exists $\tau \in \operatorname{Traces}^\omega_O(G)$ such that $\Pi[\pi \gets \tau] \models^\omega \psi'$.
    We pick $\tau' = \tau_0...\tau_{k - 1} \in \operatorname{Traces}^k_O(G)$. Clearly, $\Pi'[\pi \gets \tau'] \sqsubset \Pi[\pi \gets \tau]$ and by induction hypothesis it follows that $\Pi'[\pi \gets \tau'] \models^k \psi'$.\qedhere
  \end{itemize}
\end{proofsketch}

In the context of automated bug finding, \cref{thm:agree} is crucial as it guarantees that violations detected with respect to the bounded semantics immediately translate to violations with respect to the unbounded semantics.

The bounded semantics addresses the problem presented above: it does not ignore finite traces, nor infinite traces with only finitely many observations.
However, it still has some limitations. Importantly, $\models^{k + 1} \psi$ does not imply $\models^{k} \psi$: even if no violations are detected for trace prefixes with $k$ observations, there could still be violations for trace prefixes with $k' < k$ observations.
This is somewhat counterintuitive, in particular compared to usual bounded semantics for single trace logics such as LTL~\cite{bmc_2003}.
For example, consider the following program graph $G$

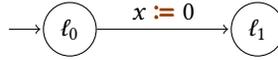
\begin{center}
  \vspace{1em}
  \begin{tikzpicture}
    \node[initial, state] (A) {$\ell_0$};
    \node[state, right of = A] (B) {$\ell_1$};
    \draw[->] (A) edge node {$\sset{x}{0}$} (B);
  \end{tikzpicture}
  \vspace{1em}
\end{center}

\noindent and the specification $\psi = \forall^\pi : \{\ell_1\}. \square(x_\pi > 0)$.
Clearly, the program should be considered to violate the specification as it sets $x$ to $0$, thus violating the invariant $x > 0$. For $k \in \{ 0, 1 \}$, we have $\not\models^k \psi$ as expected (note that $\psi$ is considered valid under the unbounded semantics!).
However, under the bounded semantics $\models^k \psi$ trivially holds for any $k > 1$ because, for such $k$'s, $\operatorname{Traces}^k_{\{\ell_1\}}(G) = \emptyset$.
Note that this problem can still occur even if we restrict ourselves to non-terminating programs. For example, for the same specification $\psi$, the following non-terminating variant of $G$ is still considered to be satisfied for $k > 1$ (because it steps through $\ell_1$ only once):

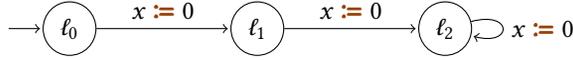
\begin{center}
  \vspace{1em}
  \begin{tikzpicture}
    \node[initial, state] (A) {$\ell_0$};
    \node[state, right of = A] (B) {$\ell_1$};
    \node[state, right of = B] (C) {$\ell_2$};
    \draw[->] (A) edge node {$\sset{x}{0}$} (B);
    \draw[->] (B) edge node {$\sset{x}{0}$} (C);
    \draw[->] (C) edge[loop right] node {$\sset{x}{0}$} (C);
  \end{tikzpicture}
  \vspace{1em}
\end{center}

To achieve a more intuitive $k$-bounded semantics, it suffices to consider bad interactions between traces with at most $k$ observations, instead of traces with exactly $k$ observations.

\begin{definition}[Upper-bounded semantics]
  Let $k \in \mathbb{N}$ and $\psi$ a specification. We define $\models^{\le k} \psi$ such that $\models^{\le k} \psi \;\iff\; \forall k' \leq k, \models^{k'} \psi$.
  \label{def:ob}
\end{definition}

\noindent
By definition, and contrary to $\models^k$, $\models^{\le k}$ enjoys the intuitive property of being monotonic.

\begin{theorem}[Monotonicity of $\models^{\le k}$]
  Let $\psi$ be a specification, and $k, k' \in \mathbb{N}$ such that $k' < k$.
  $\models^{\le k} \psi \implies \models^{\le k'} \psi$
\end{theorem}

\noindent
Further, it is easy to see that $\models^{\le k}$ still agrees with $\models^\omega$.

\begin{theorem}
  Let $\psi$ be an infinitely observable specification and $k \in \mathbb{N}$. Then $\models^\omega \psi \implies \models^{\le k} \psi$.
\end{theorem}

\section{Finding Hyperbugs by Symbolic Execution}
\label{sec:algorithm}

In the previous section, we have defined $\textrm{OHyperLTL}_\mathit{safe}$, a logic well-suited
to specify bad interactions between multiple executions of a program.
The semantics we define for $\textrm{OHyperLTL}_\mathit{safe}$ deviates from other existing relational logics such as $\textrm{OHyperLTL}$~\cite{beutner_software_2022} as it allows to consistently reason about both terminating and non-terminating executions of programs.

Even in the case of the bounded semantics of $\textrm{OHyperLTL}_\mathit{safe}$, finding a counterexample to an $\textrm{OHyperLTL}_\mathit{safe}$ formula might require inspecting an infinite number of traces. In particular, to find counterexamples to specifications of the form $\forall\pi_1\exists\pi_2\dots$, we need to identify a trace $\pi_1$ that cannot be matched with any corresponding trace $\pi_2$. In turn, this requires exhaustively exploring the set of all candidate traces $\pi_2$.
While the set of all traces of a program is usually infinite, even if their length is restricted, it is possible to efficiently
compute a (finite) symbolic representation of all traces of length $k$ using symbolic execution \cite{king_symbolic_1976}.
In this section, we take advantage of this observation to develop a symbolic encoding of the bounded semantics of $\textrm{HyperLTL}_\mathit{safe}$.
We will then use this symbolic encoding to devise an algorithm that is capable of finding counterexamples.

\subsection{Symbolic Encoding of the Bounded Semantics}

We start by defining symbolic execution for program graphs.
In the following, we let $V_\ast$ be an infinite set of unique \emph{fresh variables} over $\mathcal{T}$ such that $V_\ast \cap X = \emptyset$. Further, we suppose given a function $\operatorname{fresh}()$ that generates a new variable in $V_\ast$, different from all the other variables in context.

Symbolic execution aims to generate a symbolic encoding of sets of execution paths. Following \cite{de_boer_nature_2019} and \cite{correnson_fse_2023}, we present symbolic execution as another semantics of programs where memories, states, and traces are replaced with symbolic encodings.
In the context of symbolic execution, programs operate on a \emph{symbolic memory} $\hat{m} \in \hat{\mathcal{M}} = (\operatorname{Term}_\mathcal{T}(V_\ast))^X$ mapping program variables to symbolic expressions.
We use $\hat{m}_0$ to denote some symbolic memory that corresponds to the initial memory $m_0$, i.e., $\hat{m}_0 \in \hat{\mathcal{M}}$ satisfies $\den{\hat{m}_0(x)}_\mathcal{T}^\emptyset = m_0(x)$ for all $x \in X$. A \emph{symbolic state} is a triple $\hat{s} = \langle \ell, \varphi, \hat{m} \rangle$, where $\ell$ is a program location, $\varphi \in \operatorname{Term}_\mathcal{T}(V_\ast)$ is a quantifier-free \emph{path formula}, and $\hat{m} \in \hat{\mathcal{M}}$ is a symbolic memory.
A \emph{symbolic trace} is a finite sequence of symbolic states.
For a given symbolic trace $\hat{\tau} = \langle\ell_0, \varphi_0, \hat{m}_0\rangle \dots \langle\ell_n, \varphi_n, \hat{m}_n\rangle$, we use $\operatorname{path}(\hat{\tau}) \coloneqq \varphi_n$ to denote the accumulated path formula $\varphi_n$.
Within this model of symbolic execution, we can equip program graphs with a \emph{symbolic semantics} that defines how to compute symbolic encodings of program traces.
The symbolic semantics mimics the concrete semantics presented in \cref{def:concrete-operational-semantics}, but replaces concrete memory assignments with symbolic ones.

\begin{definition}[Symbolic operational semantics]
  \begin{mathpar}
    \inferrule[sym-assign]{ ~ }{\langle \hat{m}, \sset{x}{e} \rangle \Downarrow_\mathit{sym} \hat{m}[x \gets e[/\hat{m}]]}\qquad
    \inferrule[sym-havoc]{ ~ }{\langle \hat{m}, \srand{x}\rangle \Downarrow_\mathit{sym} \hat{m}[x \gets \operatorname{fresh()}]}\\
    \inferrule[sym-step]{ \langle \ell_1, \ell_2 \rangle \in \mathcal{E}\\\varphi_2 = \varphi_1 \sand \mathit{guard}(\ell_1, \ell_2)[/\hat{m}_1]\\\exists \rho, \rho \models_\mathcal{T} \varphi_2\\\langle \hat{m}_1, \mathit{effect}(\ell_1, \ell_2)\rangle \Downarrow_\mathit{sym} \hat{m}_2 }{G \vdash \langle \ell_1, \varphi_1, \hat{m}_1 \rangle \hookrightarrow_\mathit{sym} \langle \ell_2, \varphi_2, \hat{m}_2 \rangle}
  \end{mathpar}
\end{definition}

\begin{definition}[Symbolic traces]
  \begin{align*}
    \operatorname{SymTraces}^*(G) \coloneqq& \{ \hat{\tau} \mid \hat{\tau}_0 = \langle\ell_0, \strue, \hat{m}_0\rangle \;\land G \vdash \hat{\tau}_i \hookrightarrow_\mathit{sym} \hat{\tau}_{i + 1} \text{ for all } i < |\hat{\tau}| - 1 \}\\
    \operatorname{SymTraces}^*_O(G) \coloneqq& \{ [\hat{\tau}]_O \mid \hat{\tau} \in \operatorname{SymTraces}^*(G) \}\\
    \operatorname{SymTraces}^k_O(G) \coloneqq& \{ \hat{\tau} \in \operatorname{SymTraces}^*_O(G) \mid |\hat{\tau}| = k \}
  \end{align*}
\end{definition}

Given an assignment $\rho$ for the fresh variables generated during symbolic execution, a symbolic trace $\hat{\tau}$ can be concretized into a trace $\tau = \gamma_\rho(\hat{\tau})$ as follows:

\begin{definition}[Concretization]
  \begin{align*}
    \gamma_\rho(\langle\ell_0, \varphi_0, \hat{m}_0\rangle\dots\langle\ell_n, \varphi_n, \hat{m}_n\rangle) &\coloneqq \langle\ell_0, x \mapsto \den{\hat{m}_0(x)}_\mathcal{T}^\rho\rangle\dots\langle\ell_n, x \mapsto \den{\hat{m}_n(x)}_\mathcal{T}^\rho\rangle\\
  \gamma(\hat{\tau}) &\coloneqq \{ \gamma_\rho(\hat{\tau}) \mid \rho \models_\mathcal{T} \operatorname{path}(\hat{\tau})\}\\
  \gamma(T) &\coloneqq \bigcup\nolimits_{\hat{\tau} \in T}\gamma(\hat{\tau}), \quad \textrm{where $T$ is a set of symbolic traces}
  \end{align*}
\end{definition}

Importantly, the concretization of the set of (finite) symbolic traces is equal to the set of concrete (finite) traces. This result has been formally proven in \cite{de_boer_nature_2019} and \cite{correnson_fse_2023}.

\begin{theorem}[Equivalence of symbolic and concrete trace semantics]
  For any program graph $G$, $\gamma(\operatorname{SymTraces}^*(G)) = \operatorname{Traces}^*(G)$ holds.
  \label{thm:symex}
\end{theorem}

\noindent This result naturally transfers to the bounded trace semantics with observations.

\begin{corollary}\label{cor:equiv-sym-traces}
  Let $G$ be a program graph with program locations $\mathcal{L}$, let $O \subseteq \mathcal{L}$ be a set of observed locations, and let $k \in \mathbb{N}$. Then $\gamma(\operatorname{SymTraces}^O_k(G)) = \operatorname{Traces}^O_k(G)$.
\end{corollary}

Given a specification $\psi$ and a bound $n$, our goal is to determine whether or not $\models^{\le n} \psi$ holds.
More precisely, we are interested in finding, as quickly as possible, evidence that $\not\models^{\le n} \psi$.
To do so, we give a symbolic encoding of the bounded semantics of $\textrm{OHyperLTL}_\mathit{safe}$.
The core idea we are building towards is to evaluate $\textrm{OHyperLTL}_\mathit{safe}$ specifications on the symbolic traces of a system instead of on all the concrete traces.
This is possible because symbolic traces faithfully represent concrete ones in virtue of \cref{thm:symex}.
When evaluating an $\textrm{OHyperLTL}_\mathit{safe}$ formula with symbolic traces, universal and existential quantification are naturally replaced by conjunction and disjunction, respectively.

\begin{definition}[Symbolic encoding of $\textrm{OHyperLTL}_\mathit{safe}$]\label{def:sym-encoding} Given a bound $k \in \mathbb{N}$ and a partial map $\hat{\Pi}$ assigning symbolic traces to trace variables, we define:
  \begin{align*}
    \den{\forall^G\pi : O.\psi}^k_{\hat{\Pi}} &:= \mathcolor{symhighlight}{\bigwedge}\nolimits_{\hat{\tau} \in \operatorname{SymTraces}^O_k(G)} \sall (\operatorname{FV}(\hat{\tau})), \mathcolor{symhighlight}{\Bigl(}\operatorname{path}(\hat{\tau}) \simp \den{\psi}^k_{{\hat{\Pi}}[\pi \gets \hat{\tau}]}\mathcolor{symhighlight}{\Bigr)}\\
    \den{\exists^G\pi : O.\psi}^k_{\hat{\Pi}} &:= \mathcolor{symhighlight}{\bigvee}\nolimits_{\hat{\tau} \in \operatorname{SymTraces}^O_k(G)} \sex (\operatorname{FV}(\hat{\tau})), \mathcolor{symhighlight}{\Bigl(}\operatorname{path}(\hat{\tau}) \sand \den{\psi}^k_{{\hat{\Pi}}[\pi \gets \hat{\tau}]}\mathcolor{symhighlight}{\Bigr)}\\
    \den{\square\varphi}^k_{\hat{\Pi}} &:= \mathcolor{symhighlight}{\bigwedge}\nolimits_{0 \le i < k}\varphi[/{\hat{\Pi}}_i], \quad \textrm{where} \ {\hat{\Pi}}_i(x_\pi) := \operatorname{mem}(\hat{\Pi}(\pi)_i)(x),
  \end{align*}
  where $\operatorname{FV}(\hat{\tau}) \subset V_\ast$ is the set of all free variables that occur in a symbolic trace $\hat{\tau}$, i.e., the set of free variables in $\operatorname{path}(\hat{\tau})$ and in the image of the symbolic memories in $\hat{\tau}$.
\end{definition}

For any specification $\psi$ and bound $k$, the encoding $\den{\psi}^k_\emptyset$ as given in \cref{def:sym-encoding} is a closed first-order formula over $\mathcal{T}$ that is valid if and only if $\models^k \psi$.

\begin{theorem}[Symbolic encoding of $\models^k$]\label{thm:sound-sym-semantics}
  For any $\textrm{OHyperLTL}_\mathit{safe}$ formula $\psi$, we have $\models^k \psi$ if and only if $\models_\mathcal{T} \den{\psi}^k_\emptyset$.
\end{theorem}
\begin{proofsketch}
  We prove the stronger result \[
    \forall \Pi, \forall \hat{\Pi}, \forall \rho, \Pi \simeq_\rho \hat{\Pi} \implies (\Pi \models^k \psi \iff \rho \models_\mathcal{T} \den{\psi}^k_{\hat{\Pi}}),
  \]
  where $\simeq_\rho$ is defined as \[
    \Pi \simeq_\rho \hat{\Pi} \quad\iff\quad \operatorname{dom}(\Pi) = \operatorname{dom}(\hat{\Pi}) \;\land\; \forall \pi \in \operatorname{dom}(\Pi), \Pi(\pi) \in \gamma_\rho(\hat{\Pi}(\pi)).
  \]
  The proof goes by induction on the formula $\psi$ and directly follows from definitions and the equivalence between concrete and symbolic semantics (see \cref{thm:symex}).
\end{proofsketch}

Of course, because of this equivalence, $\den{\psi}_\emptyset^k$ exhibits the same -- perhaps surprising -- non-monotonicity that we discussed for $\models^k \psi$ in \cref{sec:ohyperltlsafe}. Nevertheless, we can use the same symbolic encoding for $\models^{\le k} \psi$, which guarantees the desired monotonicity.

\begin{corollary}[Symbolic encoding of $\models^{\le k}$]
  For any $\textrm{OHyperLTL}_\mathit{safe}$ formula $\psi$, we have $\models^{\le k} \psi$ if and only if $\models_\mathcal{T} \den{\psi}^{k'}_\emptyset$ holds for all $k' \leq k$.
\end{corollary}

\subsection{Computing Symbolic Encodings}

It is important to note that, while the set of symbolic traces of length exactly $k$ is always finite, it is not the case for observed symbolic traces $\operatorname{SymTraces}^O_k(G)$. Thus, the symbolic encoding of a $\operatorname{OHyperLTL}_\mathit{safe}$ specification cannot always be computed.
For example, consider the following program computing the factorial of some arbitrary integer $n$ and suppose we set an observation point at the end of the program (in location $\ell_4$):

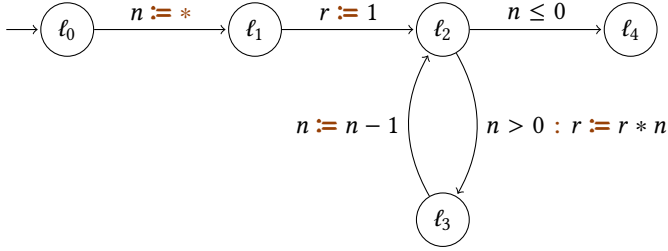
\begin{center}
  \begin{tikzpicture}
    \node[initial, state] (A) {$\ell_0$};
    \node[state, right of = A] (B) {$\ell_1$};
    \node[state, right of = B] (C) {$\ell_2$};
    \node[state, below of = C] (D) {$\ell_3$};
    \node[state, right of = C] (E) {$\ell_4$};
    \draw[->] (A) edge node {$\srand{n}$} (B);
    \draw[->] (B) edge node {$\sset{r}{1}$} (C);
    \draw[->] (C) edge[bend left] node {$n > 0 \ \syn{:} \ \sset{r}{r * n}$} (D);
    \draw[->] (D) edge[bend left] node {$\sset{n}{n - 1}$} (C);
    \draw[->] (C) edge node {$n \le 0$} (E);
  \end{tikzpicture}
\end{center}

Clearly, there exist infinitely many symbolic traces with exactly one visit to $\ell_4$, each corresponding to the execution of the program with a different input $n$.
Formally, $|\operatorname{SymTraces}^1_{ \{\ell_2\} }(G)| = \infty$, and consequently, this set cannot be computed by symbolic execution. This issue is inherent to program analysis methods based on loop unwinding: the analysis of unbounded loops never terminates.

\begin{definition}[Encodable specifications]
  Let $\psi$ be a $\textrm{HyperLTL}_\mathit{safe}$ formula and $k \in \mathit{N}$.
  We say that $\psi$ is $k$-\textit{encodable} if, for any quantified program $G$ annotated with a set of observed locations $O$ in $\psi$, $\operatorname{SymTraces}^k_O(G)$ is finite.
\end{definition}

\begin{theorem}[Finite encodings]
  Let $\psi$ be a $k$-encodable specification. Then $\den{\psi}^k_\emptyset$ is a well-defined (finite) first-order formula.
\end{theorem}

Requiring specifications to be encodable is a necessary condition to be able to effectively compute the symbolic encoding of a specification. However, it is not a sufficient criterion. Indeed, even if we restrict ourselves to the analysis of programs for which $\operatorname{SymTraces}^k_O(G)$ is finite, it does not necessarily mean that the set can be effectively computed.
This can be shown via a reduction from the halting problem. 

\begin{theorem}\label{thm:not-computable}
  The function $\operatorname{SymTraces}^k_O(G)$ is not computable for arbitrary $k \in \mathbb{N}$, program graphs $G$, and locations $O$, even if $G$ is restricted to program graphs for which $|\operatorname{SymTraces}^k_O(G)|$ is finite.
\end{theorem}
\begin{proofsketch}
  For the sake of contradiction, suppose that such an algorithm $\textproc{A}$ exists that computes the set $\operatorname{SymTraces}^k_O(G)$.
  We would then be able to construct an algorithm that solves the halting problem for an arbitrary deterministic turing machine (which is impossible).
  First, observe that any deterministic turing machine can be encoded as a program graph $P$ with a single final location $\ell_\mathit{halt}$, and whose input is chosen by setting the initial memory of $P$. Clearly, for such an encoding $P$ of a deterministic turing machine,
  $\operatorname{SymTraces}^1_{ \{\, \ell_\mathit{halt} \,\} }(P)$ is either empty (if $P$ diverges for its preloaded input) or it contains exactly one trace (if $P$ terminates for its preloaded input).
  Therefore, to decide whether $P$ terminates on the fixed input, it suffices to call $\textproc{A}$ to compute $\operatorname{SymTraces}^1_{ \{\, \ell_\mathit{halt} \,\} }(P)$ and check whether its cardinality is zero or one.
\end{proofsketch}

In spite of \cref{thm:not-computable}, we demonstrate that a sound (but necessarily incomplete) algorithm exists that computes $\operatorname{SymTraces}^k_O(G)$ by symbolic execution of $G$.
Further, we will show that this algorithm is actually complete for a large class of interesting programs we call finitely observable programs.
\cref{alg:symex} consists of the two procedures $\textsc{Extend}(\hat{\tau})$ and $\textsc{Observe}$. The first procedure takes a symbolic trace $\hat{\tau}$ of length $k$ and computes all feasible extensions of $\hat{\tau}$ of length $k + 1$.
To check the feasibility of symbolic traces, we assume the existence of a procedure $\textproc{Sat}(\varphi)$ that (semi-) decides the satisfiability of an SMT formula $\varphi$ over $\mathcal{T}$.
The procedure $\textsc{Observe}$ uses $\textsc{Extend}$ to iteratively compute $\operatorname{SymTraces}^k_O(G)$. The traces are computed by starting from the initial symbolic state $\langle \ell_0, \hat{m}_0, \mathit{true} \rangle$, and repeatedly computing all its possible extensions, until the desired number of observation points is reached for all extensions.

\algnewcommand\algorithmicswitch{\textbf{switch}}
\algnewcommand\algorithmiccase{\textbf{case}}
\algnewcommand\algorithmicassert{\texttt{assert}}
\algnewcommand\Assert[1]{\State \algorithmicassert(#1)}%
\algdef{SE}[SWITCH]{Switch}{EndSwitch}[1]{\algorithmicswitch\ #1\ \algorithmicdo}{\algorithmicend\ \algorithmicswitch}
\algdef{SE}[CASE]{Case}{EndCase}[1]{\algorithmiccase\ #1}{\algorithmicend\ \algorithmiccase}%
\algtext*{EndSwitch}
\algtext*{EndCase}

\begin{algorithm}
  \begin{algorithmic}
    \Require program graph $G$
    \Require set of observed locations $O$
    \Require number of observations $n \in \mathbb{N}$

    \bigskip
    
    \Procedure{Extend}{$\hat{\tau}$}
      \State $T \gets \emptyset$
      \State $\langle \ell, \hat{m}, \varphi \rangle \gets \hat{\tau}_{ |\hat{\tau}| - 1}$
      \ForAll{$(\ell_1, \ell_2) \in \mathcal{E}$ such that $\ell_1 = \ell$}
          \State $\varphi' \gets \varphi \sand \mathit{guard}(\ell_1, \ell_2)[/\hat{m}]$
          \If{$\textsc{Sat}(\varphi')$}
            \Switch{$\mathit{effect}(\ell_1, \ell_2)$}
            \Case{$\sset{x}{e}$}
              \State $T \gets T \cup \{ \hat{\tau} \cdot \langle \ell_2, \hat{m}[x \gets e[/\hat{m}]], \varphi' \rangle \}$
            \EndCase
            \Case{$\srand{x}$}
              \State $T \gets T \cup \{ \hat{\tau} \cdot \langle \ell_2, \hat{m}[x \gets \mathit{fresh}()], \varphi' \rangle \}$
            \EndCase
            \EndSwitch
          \EndIf
      \EndFor
      \State\Return $T$
    \EndProcedure

    \bigskip

    \Procedure{Observe}{}
      \State $T \gets \emptyset$
      \State $\mathit{todo} \gets \{ \langle \ell_0, \hat{m}_0, \texttt{true} \rangle \}$
      \While{$\mathit{todo} \ne \emptyset$}
        \State \textbf{pick} $\hat{\tau} \in \mathit{todo}$
        \State $\mathit{todo} \gets \mathit{todo} \setminus \{ \hat{\tau} \}$
        \If{$|\hat{\tau}|_O = n$}
          \State $T \gets T \cup \{ [\hat{\tau}]_O \}$
        \Else
          \State $\mathit{todo} \gets \mathit{todo} \cup \textsc{Extend}(\hat{\tau})$
        \EndIf
      \EndWhile
      \State\Return $T$
    \EndProcedure
  \end{algorithmic}
\caption{Symbolic Interpreter}
\label{alg:symex}
\end{algorithm}

\noindent
By construction, \cref{alg:symex} satisfies the following property.

\begin{theorem}[Soundness of the symbolic interpreter]
  Let $G$ be a program graph, $O$ be a set of observed locations, and $n$ a number of observations.
  Then, if $\textsc{Observe}(G, O, n)$ terminates, it returns $\operatorname{SymTraces}^n_O(G)$.
\end{theorem}

We can identify a class of \emph{finitely observable} programs for which the symbolic encoding can always be computed. In practice, we found that the vast majority of benchmarks in the relevant literature fall into this class (see \cref{sec:eval}).

\begin{definition}[Finite observability of programs]\label{def:observable}
  Let $G$ be a program with locations $\mathcal{L}$, $k \in \mathbb{N}$, and $O \subseteq \mathcal{L}$ be a set of observed locations. $G$ is said to be $(k, O)$-observable and we note $\operatorname{Observable}^O_k(G)$ if and only if there exists a bound $b \in \mathbb{N}$ such that for all finite traces $\tau$ with exactly $k$ observations, there exists a prefix of $\tau$ of length at most $b$ with the same $k$ observations. Formally, \[
    \exists b \in \mathbb{N}, \forall \tau \in \operatorname{Traces}^*(G), \; |\tau|_O = k \implies \exists \tau' \preceq \tau, |\tau'|_O = k \wedge |\tau'| < b,
  \]
  where $\tau' \preceq \tau \iff \forall i \in \{\, 0, \dots, |\tau'| - 1 \,\},\; \tau'_i = \tau_i$ for a pair of finite traces $\tau = \tau_0 \dots \tau_n, \;\tau' = \tau'_0 \dots \tau'_{n'} \in \operatorname{Traces}^*(G)$.
\end{definition}

\begin{theorem}[Relative completeness of the symbolic interpreter]
  Let $G$ be a program graph, $O$ a set of observed locations, $n \in \mathbb{N}$ a number of observations, and suppose that $\mathcal{T}$ is a decidable theory.
  If $G$ is $(n, O)$-observable, then $\textsc{Observe}(G, O, n)$ terminates and returns $\operatorname{SymTraces}^n_O(G)$.
\end{theorem}

\begin{corollary}[Computability of the encoding]
  Let $\psi$ be an $\textrm{OHyperLTL}_\mathit{safe}$ formula and let $k \in \mathbb{N}$ such that for all quantification patterns $Q^G\pi:O$ in $\psi$, $\operatorname{Observable}^O_k(G)$ holds.
  Then, the symbolic encoding $\den{\psi}^k_\emptyset$ can be effectively computed.
\end{corollary}

\subsection{Automated Bug-Finding Algorithms}
\label{sec:algorithm-subsection}

The above definitions and theorems are sufficient for us to formulate a first algorithm that utilizes the symbolic encoding $\den{\psi}^k_\emptyset$ to determine whether or not $\models^{\le k} \psi$ holds. This approach is depicted in \cref{alg:naive}.

\begin{algorithm}
  \begin{algorithmic}
    \Require $\textrm{OHyperLTL}_\mathit{safe}$ formula $\psi$
    \Require $n \in \mathbb{N}$\\
    \Procedure{Naive}{}
      \ForAll{$k$ from $1$ to $n$}
        \If{$\Call{Sat}{\snot\den{\psi}^k_\emptyset}$}
          \State\Return \enquote{bug found}
        \EndIf
      \EndFor
    \EndProcedure
  \end{algorithmic}
\caption{Naive bug-finding algorithm}
\label{alg:naive}
\end{algorithm}

\noindent The correctness of \cref{alg:naive} is formally stated as follows.

\begin{theorem}[Soundness of \cref{alg:naive}]
  For any $\textrm{OHyperLTL}_\mathit{safe}$ formula $\psi$ and $n \in \mathbb{N}$, if $\textproc{Naive}(\psi, n)$ reports a specification violation during some iteration $k$, then $\not\models^{\le n}~\psi$.
\end{theorem}

If the specification $\psi$ is not finitely observable, \cref{alg:naive} might fail to compute the symbolic encoding and diverge.
If we require input specifications to be finitely observable, \cref{alg:naive} is complete with respect to bug finding.
Note that this completeness result is relative to the decidability of the underlying first-order theory.

\begin{theorem}[Relative completeness of \cref{alg:naive}]\label{thm:naive-rel-compl}
  Let $\psi$ be a finitely observable specification, $n \in \mathbb{N},$ and suppose that $\mathcal{T}$ is a decidable theory.
  If $\not\models^{\le n} \psi$, then $\textproc{Naive}(\psi, n)$ reports a violation.
\end{theorem}

Nevertheless, \cref{alg:naive} is unsatisfactory.
Perhaps most importantly, our goal is not only to prove $\not\models^{\le n} \psi$, but also to produce a witness of this violation.
Because $\den{\psi}_\emptyset^k$ is a closed formula, $\textproc{Sat}(\snot\den{\psi}_\emptyset^k)$ -- and thus \cref{alg:naive} -- does not produce any (non-empty) model.
Additionally, the size of the symbolic encoding $\snot\den{\psi}_\emptyset^k$ grows rapidly, and the search for relevant traces generally cannot be guided as the entire problem is now encoded as a single SMT query (for each value of $k$).
Lastly, \cref{alg:naive} is not well-suited for parallelization, which is a desirable property for almost any computationally extensive algorithm due to the parallel architecture of modern processors.

While \cref{alg:naive} works for arbitrary $\textrm{OHyperLTL}_\mathit{safe}$ formulas, we are particularly interested in the class of $\forall\exists$ hyperproperties.
Therefore, we now assume a specification of the form $\psi = \forall^{G_1}\pi_1 : O_1.\exists^{G_2}\pi_2 : O_2.\square \varphi$.
Given such a specification, we can now improve the search for a counterexample in multiple ways.
First, symbolic traces of $G_1$ can be processed one by one.
Instead of generating one big conjunctive query, we try to find a counterexample for every symbolic trace of $G_1$ independently.
More precisely, for every symbolic trace in $G_1$, we try to prove that there exists no matching trace in $G_2$.
This drastically reduces the size of SMT queries while also allowing more flexibility in the selection of symbolic paths of $G_1$ that are considered.
This approach is depicted in \cref{alg:smart}.

\begin{algorithm}
  \begin{algorithmic}
    \Require  $\textrm{OHyperLTL}_\mathit{safe}$ formula $\psi = \forall^{G_1}\pi_1 : O_1.\exists^{G_2}\pi_2 : O_2.\square \varphi$
    \Require $n \in \mathbb{N}$\\
    \Procedure{Lazy}{}
      \ForAll{$k$ from $1$ to $n$}
        \ForAll{$\hat{\tau}_1 \in \operatorname{SymTraces}^{O_1}_k(G_1)$}
          \State $C_1 \gets \operatorname{path}(\hat{\tau}_1)$
          \State $C_2 \gets \snot\den{\exists^{G_2} \pi_2 \colon O_2 . \square\varphi}^k_{[\pi_1 \gets \hat{\tau}_1]}$
          \State $C \gets C_1 \sand C_2$
          \If{$\Call{Sat}{C}$}
            \State\Return $\langle\hat{\tau}_1, \operatorname{model}(C), C_2\rangle$
          \EndIf
        \EndFor
      \EndFor
    \EndProcedure
  \end{algorithmic}
\caption{\enquote{Lazy} bug-finding algorithm with counterexamples}
\label{alg:smart}
\end{algorithm}

Additionally, \cref{alg:smart} now permits recovery of a concrete trace $\tau_1$ of $G_1$ such that no corresponding concrete trace $\tau_2$ of $G_2$ can be found.
By construction, the free variables in the query $C$ are precisely the free variables in the universally quantified trace $\hat{\tau}_1$.
In other words, $\rho \coloneqq \operatorname{model}(C)$ can be used to construct a concrete trace $\tau_1 = \gamma_{\rho}(\hat{\tau}_1)$ for $\pi_1$ that witnesses the violation of the specification. The query $C_2$ encodes the impossibility of finding a matching trace in $G_2$ and acts as an \emph{explanation} as to why $\tau_1$ constitutes a counterexample of $\psi$.

The correctness of this algorithm follows from the described construction of $\rho$ and $\sigma_1$ and \cref{thm:sound-sym-semantics}.

\begin{theorem}[Soundness of \cref{alg:smart}]
  For any bound $n \in \mathbb{N}$, and any $\textrm{OHyperLTL}_\mathit{safe}$ formula $\psi = \forall^{G_1} \pi_1 \colon O_1 . \exists^{G_2} \pi_2 \colon O_2 . \square\varphi$, if $\textproc{Lazy}(\psi, n)$ reports a specification violation then $\not\models^{\le k}~\psi$.
\end{theorem}

The relative completeness theorem (\cref{thm:naive-rel-compl}) obtained for \cref{alg:naive} can also be attained for \cref{alg:smart}.

\begin{theorem}[Relative completeness of \cref{alg:smart}]
  Let $\psi$ be an $n$-observable specification of the form $\psi = \forall^{G_1} \pi_1 \colon O_1 . \exists^{G_2} \pi_2 \colon O_2 . \square\varphi$, and suppose that $\mathcal{T}$ is a decidable theory.
  If $\not\models^{\le n} \psi$, then $\textproc{Lazy}(\psi)$ reports a violation.
\end{theorem}

With this in mind, \cref{alg:smart} can even be applied to some instances for which $\operatorname{SymTraces}_k^{O_1}(G_1)$ is infinite, since traces are processed one by one.
Indeed, a counterexample might still be found by looking only at a few of the infinitely many traces in $\operatorname{SymTraces}_k^{O_1}(G_1)$.
The likelihood of locating any existing specification violation within a finite amount of time can further be affected by adjusting the order in which one iterates over $\operatorname{SymTraces}_k^{O_1}(G_1)$.

\cref{alg:smart} -- and its inner loop, in particular -- is well-suited for parallelization:
aside from sharing data as part of further optimizations, the iterations of the inner loop are entirely independent of one another and thus can be executed concurrently.

\subsection{Generalization}
\label{sec:algorithm-generalization}

\cref{alg:smart} assumes that the input formula $\psi$ is of the form $\forall^{G_1} \pi_1 \colon O_1 \exists^{G_2} \pi_2 \colon O_2 . \square\varphi$, which permits the specification of important hyperproperties such as refinement.
Nevertheless, prominent security properties, such as generalized non-interference~\cite{mccullough_noninterference_1988} and delimited information release~\cite{sabelfeld_model_2004}, fall in the class of $\forall^+\exists^+$ hyperproperties, where sequences of one or more of the same quantifier are allowed.
Our approach can be generalized to such hyperproperties of this form in a straightforward manner.
For any quantifier prefix of the form $\forall^{G^\forall_1}. \ldots \forall^{G^\forall_m} \exists^{G^\exists_1} \ldots \exists^{G^\exists_n}.$, we can use either self-composition~\cite{barthe_secure_2004} or a product program construction~\cite{barthe_relational_2011} to construct two programs $G^\forall = G^\forall_1 \otimes \ldots \otimes G^\forall_m$ and $G^\exists = G^\exists_1 \otimes \ldots \otimes G^\exists_n$ such that $\operatorname{Traces}^*(G^\forall) \simeq \operatorname{Traces}^*(G^\forall_1) \times \ldots \times \operatorname{Traces}^*(G^\forall_m)$ and $\operatorname{Traces}^*(G^\exists) \simeq \operatorname{Traces}^*(G^\exists_1) \times \ldots \times \operatorname{Traces}^*(G^\exists_n)$.
We can then replace the quantifier prefix with $\forall^{G^\forall} \pi_1. \exists^{G^\exists} \pi_2$ and rename the variables in the body of the formula accordingly to obtain an equivalent OHyperLTL specification with only two quantifiers.

Such product constructions are widely used for the formal verification of $k$-safety hyperproperties~\cite{barthe_secure_2004, yang_lazy_2018, shemer_property_2019, farzan_automated_2019}.
A straightforward way to generate the product of $n$ programs is to execute the $n$ programs in a lock-step fashion.
However, lock-step products often complicate the verification of hyperproperties as they require to find intricate relational invariants~\cite{shemer_property_2019, farzan_automated_2019}.
In turn, one of the main challenges in the verification of hyperproperties based on product constructions is to find an appropriate interleaving of the original programs such that the verification is easy (e.g., such that there exists a simple invariance argument)~\cite{farzan_automated_2019}.
In the context of this work, we are not concerned with formal verification but rather with automated bug-finding. Therefore, finding interleavings leading to easier invariants it not a concern. However, a naive lock-step product construction would still not work. Indeed, the hyperproperties we are focusing on require to synchronize the different traces at specific observation points; this requires additional care. Instead of using a naive lock-step product, we use an \textit{asynchronous product} construction. The idea is to synchronize the execution of the different programs whenever they reach an observation point.
We give a detailed explanation of this construction in the next subsection.

\subsection{Asynchronous Product Construction}

We describe our product construction for pairs of program graphs. We consider two program graphs $G_1$ and $G_2$ with locations $\mathcal{L}_1$ and $\mathcal{L}_2$, respectively, as well as two sets of observed location $O_1 = \{ o^1_1, \ldots, o^1_{|O_1|} \} \subseteq \mathcal{L}_1$ and $O_2 = \{ o^2_1, \ldots, o^2_{|O_2|} \} \subseteq \mathcal{L}_2$. Without loss of generality, we assume that $G_1$ and $G_2$ do not share any program variables.
Our goal is to construct a product program graph $G_{1 \otimes 2}$ that mimics the simultaneous execution of $G_1$ and $G_2$. To compute the product of more than two program graphs, this construction can be applied repeatedly.

The high-level intuition of our construction is to start by executing (a copy of) $G_1$ until it reaches any observation point in $O_1$, then execute (a copy of) $G_2$ until it reaches any observation point in $O_2$. Once both programs reached a first observation point, the product program resumes the execution of $G_1$ until the next observation point in $O_1$, then transfers again the control to $G_2$ until the next observation point in $O_2$, etc.
The end of every such \textit{round} of execution is marked as an observation point in the final product program. We use multiple copies of $G_1$ and $G_2$ to memorize where to jump next during transitions from $G_1$ to $G_2$ and from $G_2$ to $G_1$.

To efficiently construct the product program, we perform the following sequence of transformations on the original program graphs $G_1$ and $G_2$. \begin{enumerate}[leftmargin=*]
  \item First, for every observed location $o^1_i$ in $G_1$, we duplicate this location by creating a fresh location $r^1_i$ which we call the \textit{re-entry point} of $o^1_i$. All outgoing edges of $o^1_i$ are deleted, and $r^1_i$ inherits these edges instead. We perform the same operation on observed locations of $G_2$.
  \item Second, we create $|O_2| + 1$ copies of $G_1$ (after the integration of \textit{re-entry} points). We name these copies $G_1(0), G_1(1), \ldots, G_1(|O_2|)$. Similarly, we create $|O_1|$ copies $G_2(1), \ldots, G_2(|O_1|)$ of $G_2$.
  The intuition is that each copy of $G_1$ (resp. $G_2$) tracks which was the last observed location visited by $G_2$ (resp. $G_1$). The initial location of the product program $G_{1 \otimes 2}$ is set to be the initial location of the first copy of $G_1$, i.e., of $G_1(0)$.
  \item Next, we insert effect-free transitions between every observed location and some appropriate re-entry point.
  More precisely, we redirect any observed location $o^1_i$ in $G_1(0)$ to the initial location of $G_2$ in $G_2(i)$.
  For $j \geq 1$, reaching $o^1_i$ in $G_1(j)$ redirects to the re-entry point $r^2_j$ in $G_2(i)$.
  Symmetrically, reaching $o^2_i$ in $G_2(j)$ for $j \geq 1$ redirects to the re-entry point $r^1_j$ in $G_1(i)$.
  \item Finally, we mark every observed location $o^2_i$ in any copy of $G_2$ as an observed location. We note $O_{1 \otimes 2}$ the corresponding set of locations. The resulting product program is such that $\mathit{Traces}^k_{O_{1 \otimes 2}}(G_{1 \otimes 2}) = \{ \tau_1 \uplus \tau_2 \mid (\tau_1, \tau_2) \in \mathit{Traces}^k_{O_1}(G_1) \times \mathit{Traces}^k_{O_2}(G_2) \}$, where $\uplus$ denotes the point-wise union of states along two traces.
\end{enumerate}

To better illustrate how the construction works, consider an example consisting of two program graphs $G_1$ and $G_2$ that each have a single observed location.
Schematically, these two program graphs can be represented as follows: 
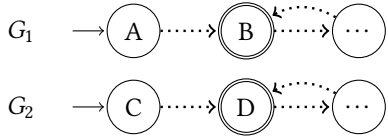
\begin{center}
  \begin{tikzpicture}
    \node (P1) {$\mathit{G}_1$};
    \node[initial, state, right of = P1, xshift=-1cm] (A1) {A};
    \node[accepting, state, right of = A1, xshift=-1cm] (B1) {B};
    \node[state, right of = B1, xshift=-1cm] (C1) {\ldots};

    \node[below = of P1, yshift=1.5cm] (P2) {$\mathit{G}_2$};
    \node[initial, state, right of = P2, xshift=-1cm] (A2) {C};
    \node[accepting, state, right of = A2, xshift=-1cm] (B2) {D};
    \node[state, right of = B2, xshift=-1cm] (C2) {\ldots};

    \draw[->, dotted, line width=0.9pt] (A1) edge (B1);
    \draw[->, dotted, line width=0.9pt] (B1) edge (C1);
    \draw[->, dotted, line width=0.9pt] (C1) edge[bend right] (B1);
    
    \draw[->, dotted, line width=0.9pt] (A2) edge (B2);
    \draw[->, dotted, line width=0.9pt] (B2) edge (C2);
    \draw[->, dotted, line width=0.9pt] (C2) edge[bend right] (B2);
  \end{tikzpicture}
\end{center}

After their initial location, each program executes some arbitrary code and eventually reaches its observed location, which is represented by a double circle. Afterwards, the execution continues with some arbitrary code and might eventually cycle back to the observation point.
The figure below decomposes the different steps of constructing the product of two such program graphs.
First, on the left, we introduce the re-entry points and we create copies of $G_1$ and $G_2$.
Then, on the right, we connect observed locations with their corresponding re-entry points, and we mark the initial state and the new observation point.

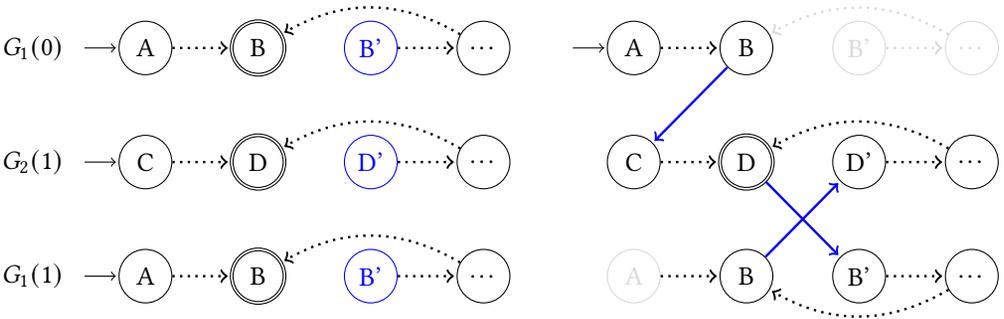
\begin{center}
  \begin{tikzpicture}
    \node (P1) {$\mathit{G}_1(0)$};
    \node[initial, state, right of = P1, xshift=-1cm] (A1) {A};
    \node[accepting, state, right of = A1, xshift=-1cm] (B1) {B};
    \node[state, right of = B1, xshift=-1cm, blue] (C1) {B'};
    \node[state, right of = C1, xshift=-1cm] (D1) {\ldots};
    
    \node[below = of P1, yshift=1cm] (P2) {$\mathit{G}_2(1)$};
    \node[initial, state, right of = P2, xshift=-1cm] (A2) {C};
    \node[accepting, state, right of = A2, xshift=-1cm] (B2) {D};
    \node[state, right of = B2, xshift=-1cm, blue] (C2) {D'};
    \node[state, right of = C2, xshift=-1cm] (D2) {\ldots};

    \node[below = of P2, yshift=1cm] (P3) {$\mathit{G}_1(1)$};
    \node[initial, state, below = of A2, yshift=1cm] (A3) {A};
    \node[accepting, state, right of = A3, xshift=-1cm] (B3) {B};
    \node[state, right of = B3, xshift=-1cm, blue] (C3) {B'};
    \node[state, right of = C3, xshift=-1cm] (D3) {\ldots};

    \draw[->, dotted, line width=0.9pt] (A1) edge (B1);
    \draw[->, dotted, line width=0.9pt] (C1) edge (D1);
    \draw[->, dotted, line width=0.9pt] (D1) edge[bend right] (B1);
    
    \draw[->, dotted, line width=0.9pt] (A2) edge (B2);
    \draw[->, dotted, line width=0.9pt] (C2) edge (D2);
    \draw[->, dotted, line width=0.9pt] (D2) edge[bend right] (B2);
    
    \draw[->, dotted, line width=0.9pt] (A3) edge (B3);
    \draw[->, dotted, line width=0.9pt] (C3) edge (D3);
    \draw[->, dotted, line width=0.9pt] (D3) edge[bend right] (B3);
  
    \node[initial, state, right of = D1, xshift=-.5cm] (A1') {A};
    \node[state, right of = A1', xshift=-1cm] (B1') {B};
    \node[state, right of = B1', xshift=-1cm, gray!30] (C1') {B'};
    \node[state, right of = C1', xshift=-1cm, gray!30] (D1') {\ldots};

    \node[state, right of = D2, xshift=-.5cm] (A2') {C};
    \node[accepting, state, right of = A2', xshift=-1cm] (B2') {D};
    \node[state, right of = B2', xshift=-1cm] (C2') {D'};
    \node[state, right of = C2', xshift=-1cm] (D2') {\ldots};

    \node[state, right = of D3, xshift=-.5cm, gray!30] (A3') {A};
    \node[state, right of = A3', xshift=-1cm] (B3') {B};
    \node[state, right of = B3', xshift=-1cm] (C3') {B'};
    \node[state, right of = C3', xshift=-1cm] (D3') {\ldots};

    \draw[->, dotted, line width=0.9pt] (A1') edge (B1');
    \draw[->, dotted, line width=0.9pt, gray, gray!30] (C1') edge (D1');
    \draw[->, dotted, line width=0.9pt, gray, gray!30] (D1') edge[bend right] (B1');
    
    \draw[->, dotted, line width=0.9pt] (A2') edge (B2');
    \draw[->, dotted, line width=0.9pt] (C2') edge (D2');
    \draw[->, dotted, line width=0.9pt] (D2') edge[bend right] (B2');
    
    \draw[->, dotted, line width=0.9pt] (A3') edge (B3');
    \draw[->, dotted, line width=0.9pt] (C3') edge (D3');
    \draw[->, dotted, line width=0.9pt] (D3') edge[bend left] (B3');

    \draw[->, blue, line width=0.9pt] (B1') edge (A2');
    \draw[->, blue, line width=0.9pt] (B2') edge (C3');
    \draw[->, blue, line width=0.9pt] (B3') edge (C2');
  \end{tikzpicture}
\end{center}

\section{Implementation and Evaluation}
\label{sec:eval}

To evaluate the effectiveness of our approach, we implemented the lazy $\forall\exists$ bug-finding algorithm (\cref{alg:smart}) in a prototype tool. Our implementation accepts structured programs expressed in a simple imperative programming language with loops, conditionals, and non-deterministic assignments, as well as an \verb|observe| instruction to annotate programs with observation points.
An input file for our tool is composed of a list of program definitions, and a $\textrm{OHyperLTL}_\mathit{safe}$ formula with a prefix of the form $\forall^*\exists^*$ quantifying over the traces of the defined programs.
Input programs are automatically translated into an equivalent program graph (as established in \cref{def:program-graph}). If the specification contains more than one universal or existential quantifier, we use a product construction (on the program graphs) to transform it into a equivalent $\forall^1\exists^1$ specification as described in \cref{sec:algorithm-generalization}.
Symbolic traces of program graphs are then computed by a custom symbolic execution engine. To check the feasibility of paths condition, and the satisfiability of the symbolic encodings, we use the Yices SMT solver \cite{dutertre_yices_2014}.

We evaluated our prototype against a diverse set of examples, many of which we drew from related work~\cite{dickerson_rhle_2022}.
These examples include several instances of refinement, generalized non-interference~\cite{mccullough_noninterference_1988}, and delimited information release~\cite{sabelfeld_model_2004}, as well as various other problems.
The main question we try to answer with this evaluation are the following:

\begin{itemize}
  \item[(Q1)] Can our approach effectively find counter-examples to relevant $\forall\exists$ hyperproperties without any expert assistance?
  \item[(Q2)] Is our approach able to report counterexamples in a reasonable time?
  \item[(Q3)] How does our approach perform when the number of paths to check and/or number of observation points grow?
\end{itemize}

To answer these questions, we evaluated our tool against the benchmarks of ORHLE~\cite{dickerson_rhle_2022}, a state-of-the-art deductive verifier for $\forall\exists$ hyperproperties of sequential programs. We additionally developed a new set of benchmarks specifically designed to stress-test our tool and measure its ability to scale when the number of program paths to explore grows.
These new benchmarks consist of reactive programs and require to unwind several observation points in order to detect a hyperproperty violation. This type of analysis falls out of the scope of ORHLE. To the best of our knowledge, our tool is the only bug-finding tool for $\forall\exists$ hyperproperties of such infinite-state reactive systems. In order to still have a comparison point, we compared the performance of our tool against HyperQB~\cite{groote_bounded_2021} and HyHorn~\cite{verification_as_chc}. HyperQB is a bounded model-checker for HyperLTL. It targets finite-state reactive systems and reduces the bug-finding problem to QBF.
HyHorn is a very recent automated verifier for $\forall\exists$ hyperproperties. It targets the verification of infinite-state reactive systems against hyperproperties expressed in a logic similar to $\textrm{OHyperLTL}_\mathit{safe}$. Importantly, like ORHLE, HyHorn is not fully automatic and it typically requires a user-supplied predicate abstraction~\cite{jhala_predicate_2018} to be able to prove certain instances. However, it also features a fully automatic "concrete" mode that does not require any guidance
and can also detect violations in some cases.
While HyperQB and HyHorn tackle slightly different problems than our tool (HyperQB focuses on finite-state systems, and HyHorn focuses on verification rather than bug-finding), we believe that these two tools were the most relevant comparison points at the time of writing.

The experimental evaluation was conducted inside Docker containers on Linux 5.10.0, on a system equipped with 1024 GiB of memory and two Intel Xeon Silver 4314 processors, each of which has 32 logical CPU cores. Note that our prototype implementation is entirely single-threaded and thus, unlike other tools, does not benefit from the large number of available CPU cores.

\subsection{Comparison with ORHLE}

As an initial evaluation of our tool, we reused benchmarks from the tool ORHLE~\cite{dickerson_rhle_2022}.
It contains a total of 35 programs and associated hyperproperties including generalized non-interference~\cite{mccullough_noninterference_1988} and refinement, which are expressed as relational pre and post-conditions.
Of these, 30 can be expressed in our input programming language (the remaining five use programming constructs that our prototype does not support). Among the 30 remaining programs, 15 contain hyperproperty violations (the others satisfy their specifications). We translated these 15 example programs into our input programming language, and expressed their associated specifications in $\textrm{OHyperLTL}_\mathit{safe}$. Importantly, since these programs are sequential, there is only one observation point at the end of each program, which corresponds to the point at which the post-condition needs to be verified.

The results of running our tool on these 15 benchmarks are reported in \cref {tab:perf-orhle}. The table recalls the type of hyperproperty being tested, the name of the program, whether the program is finitely observable or not (\textbf{FO}), and the number of different combinations of paths that our algorithm checked (\textbf{\# Combinations}). The runtime for each example program is expressed in seconds, and corresponds to the median runtime over 100 executions of our tool and ORHLE.

\newcommand{\ok}{\ding{51}}
\newcommand{\ko}{\ding{55}}

\begin{table}[ht!]
  \caption{Runtime of counterexample detection in benchmarks derived from \cite{dickerson_rhle_2022}.}
  \centering\fontsize{7}{8}\selectfont
  \def\arraystretch{1.2}
  \setlength\tabcolsep{1.2mm}
  \include{oopsla-auto-generated/orhle-benchmarks-table}
  \label{tab:perf-orhle}
\end{table}

As shown in \cref{tab:perf-orhle}, our algorithm produces a counterexample for 14 out of the 15 examples, in less than a second, and without any expert intervention. In particular, contrary to ORHLE~\cite{dickerson_rhle_2022}, our tool does not require loops to be annotated with invariants.
Our tool was not able to find the violation for one of the 15 benchmarks. This program can diverge without ever reaching its observation point, and thus falls out of the class of programs supported by our approach (in particular, it is not finitely observable).

Even though \cref{alg:smart} would naturally benefit from parallelization, our prototype implementation does not utilize this. ORHLE, on the other hand, uses multiple CPU cores and benefits from the available hardware parallelism. Nevertheless, with the exception of \texttt{loop-nonrefinement}, our tool finds specification violations significantly faster than ORHLE in these benchmarks.

\subsection{Custom Benchmarks}
\label{sec:custom-benchmarks}

In the previous set of benchmarks, the number of combination of paths that need to be checked is rather small (at most 48).
Further, because of the sequential nature of the programs being analyzed, the hyperproperties are checked only on the final state of each quantified program.
Thus, these benchmarks are not exploiting the full potential of $\textrm{OHyperLTL}_\mathit{safe}$.
To further demonstrate the effectiveness of our algorithm, both for an increasing number of observation points, and for large numbers of symbolic paths, we designed a second set of benchmarks. For this set of benchmarks, counterexamples exist but require to check relation at multiple observation points, or to check thousands of combinations of paths.
The runtime of our implementation for these benchmarks is depicted in \cref{tab:perf-our-own}.

\begin{table}[ht!]
  \caption{Runtime of counterexample detection for test instances. All input programs are observable, and none of the specifications are valid.}
  \centering\fontsize{7}{8}\selectfont
  \def\arraystretch{1.2}
  \setlength\tabcolsep{1.2mm}
  \include{oopsla-auto-generated/our-benchmarks-table}
  \label{tab:perf-our-own}
\end{table}

To highlight important aspects of this set of benchmarks, we briefly discuss the benchmark \texttt{escalating}, which is one of the thirteen instances.
It consists of the two programs \textsc{Escalating} and \textsc{Limit} and a $\forall\exists$ specification that universally quantifies over traces of the first program and existentially quantifies over traces of the second program.
The programs are synchronized at the beginning of their respective loops, i.e., their state is observed whenever they enter a loop iteration.

\begin{center}
\vspace{1em}
\begin{minipage}{.4\textwidth}
Program \textsc{Escalating}:
\begin{center}
\begin{tabular}{l}
  $\mathit{x} \gets 0$\\
  $\mathit{y} \gets 0$\\
  \textbf{loop} \\
  \quad\textbf{if} $x \equiv 0 \pmod 2$ \textbf{then}\\
  \quad\quad $y \gets y + 1$\\
  \quad \textbf{else}\\
  \quad\quad $y \gets y + x$\\
  \quad \textbf{havoc} $s \in \{\, 1, 2 \,\}$\\
  \quad $x \gets x + s$
\end{tabular}
\end{center}
\end{minipage}%
\begin{minipage}{.4\textwidth}
Program \textsc{Limit}:
\begin{center}
\begin{tabular}{l}
  $\mathit{max} \gets 15$\\
  \textbf{loop} \\
  \quad \textbf{either}\\
  \quad\quad $max \gets max + 1$\\
  \quad \textbf{or}\\
  \quad\quad\textbf{skip}\\
  \quad\\
  \quad\\
  \quad
\end{tabular}
\end{center}
\end{minipage}
\begin{center}
Specification\\
\fbox{$\forall \pi_1.\, \exists \pi_2.\, \square (y_{\pi_1} \leq max_{\pi_2})$}
\end{center}
\vspace{1em}
\end{center}

The hyperproperty does not hold because, eventually, the value of $y$ in the left program will exceed the value of $max$ in the right program.
To arrive at this conclusion, the algorithm needs to explore an increasing number of iterations and thus has to consider an exponential number of paths in the right program (\textsc{Limit}).
The left program assigns a nondeterministically chosen value to $s$ and thus also to $x$ in each iteration, and the choice of the assigned value affects what branch will be taken in the next loop iteration.
Therefore, in the left program, the number of feasible paths also grows exponentially in the number of loop iterations.
As shown in \cref{tab:perf-our-own}, our approach finds the violation after exploring seven iterations of the loop and hundreds of different path combinations within about 0.3 seconds.

\subsection{Case Study: \textsc{Escalating}}

In \cref{sec:custom-benchmarks}, we briefly explained the benchmark \texttt{escalating}, which we designed to have a number of paths that grows exponentially with the exploration depth, as well as a large state space.
We further demonstrate the challenges posed by this kind of input program by parameterizing the benchmark. By varying the initial value of the program variable $max$ in the existentially quantified program \textsc{Limit}, we can effectively adjust the difficulty of the benchmark.
Smaller initial values lead to a violation of the hyperproperty within a few loop iterations, whereas larger initial values drastically increase both the number of loop iterations after which the violation occurs as well as the size of the state space (see \cref{tab:escalating-params}).

\newcommand{\narrowdots}{\scalebox{0.7}[1.0]{\dots}}

\begin{table}
\caption{Properties of the parameterized \texttt{escalating} benchmark}
{\centering\fontsize{7}{8}\selectfont\begin{tabular}{c|c|c|c|c|c|c|c|c|c|c|c|c|c|c|c|c|c|c|c|c}
Initial $max$ & 0 & 1 & 2 & \narrowdots & 5 & 6 & \narrowdots & 11 & 12 & \narrowdots & 19 & 20 & \narrowdots & 29 & 30 & \narrowdots & 41 & 42 & \narrowdots & 55 \\\hline
Analysis depth & 4 & 4 & 5 & \narrowdots & 5 & 6 & \narrowdots & 6 & 7 & \narrowdots & 7 & 8 & \narrowdots & 8 & 9 & \narrowdots & 9 & 10 & \narrowdots & 10 \\
Max. of $x$ & 6 & 6 & 8 & \narrowdots & 8 & 10 & \narrowdots & 10 & 12 & \narrowdots & 12 & 14 & \narrowdots & 14 & 16 & \narrowdots & 16 & 18 & \narrowdots & 18 \\
Max. of $y$ & 5 & 5 & 10 & \narrowdots & 10 & 17 & \narrowdots & 17 & 26 & \narrowdots & 26 & 37 & \narrowdots & 37 & 50 & \narrowdots & 50 & 65 & \narrowdots & 65 \\
Max. of $max$ & 3 & 4 & 6 & \narrowdots & 9 & 11 & \narrowdots & 16 & 18 & \narrowdots & 25 & 27 & \narrowdots & 36 & 38 & \narrowdots & 49 & 51 & \narrowdots & 64 \\
\end{tabular}}
\label{tab:escalating-params}
\end{table}

\begin{figure}
\begin{tikzpicture}
\begin{semilogyaxis}[
  width=\textwidth,
  height = 6cm,
  xmin = -1,
  xmax = 56,
  legend pos = south east,
  legend cell align = left,
  xlabel = {Initial value of $\mathit{max}$ in the program $\textsc{Limit}$},
  ylabel = {Median runtime in seconds}
]
\addplot+[only marks, mark size = 1.5pt, mark = asterisk] table[x = p, y = hyena0]{oopsla-auto-generated/escalating_graph_data.dat};
\addlegendentry{Our tool}
\addplot+[only marks, mark size = 1.5pt, mark = x] table[x = p, y = hyperqb]{oopsla-auto-generated/escalating_graph_data.dat};
\addlegendentry{HyperQB}
\addplot+[only marks, mark size = 1.8pt, mark = diamond] table[x = p, y = hyhorn]{oopsla-auto-generated/escalating_graph_data.dat};
\addlegendentry{HyHorn}
\end{semilogyaxis}
\end{tikzpicture}
\caption{Median runtimes of our approach, HyperQB~\cite{groote_bounded_2021}, and HyHorn~\cite{verification_as_chc} on instances of \texttt{escalating}. Missing data points indicate timeouts after 30 minutes. The vertical axis is logarithmic.}
\Description{Median runtimes of our approach, HyperQB~\cite{groote_bounded_2021}, and HyHorn~\cite{verification_as_chc} on instances of \texttt{escalating}. Missing data points indicate timeouts after 30 minutes.}
\label{fig:escalating-graph}
\end{figure}
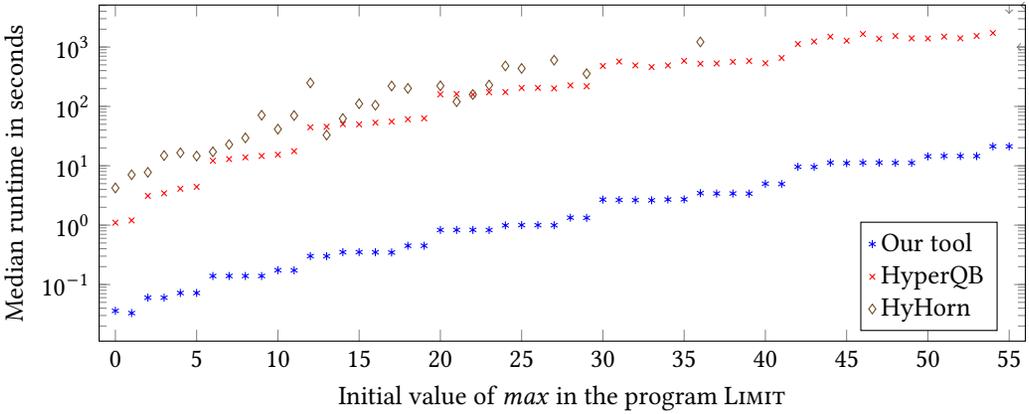

At the time of writing, there is no particularly common or standardized input format for tools that verify or refute $\forall\exists$ hyperproperties.
The vast differences across tools' capabilities and input formats make it difficult to use the same benchmarks across tools.
In spite of this, we encoded the same parameterized benchmark \texttt{escalating} not only for our own prototype tool, which consumes programs written in an imperative programming language, but also for HyperQB~\cite{groote_bounded_2021}, which consumes NuSMV files, and for HyHorn~\cite{verification_as_chc}, which uses a custom input format to model infinite-state transition systems.
This allows us to measure the runtime of all three tools on the various instances of this parameterized benchmark.

For HyperQB, we explicitly use the minimum unwinding depth that is necessary to discover the specification violation. Unlike our approach and HyHorn, HyperQB only operates on finite-state systems, which is why a general comparison is difficult.
However, as shown in \cref{tab:escalating-params}, the program's state space up until the point of the specification violation is finite, and hence, we can encode every instance as a finite-state NuSMV file for HyperQB.

For our own tool, we let it discover the appropriate depth automatically, which is its default strategy and which is compatible with the semantics of $\models^{\le k}$ (see \cref{def:ob}).
As both our tool and HyperQB are fully automatic tools, we also ran HyHorn in fully automated mode (i.e., without a user-supplied predicate abstraction). We note that even in this abstraction-less mode, HyHorn typically outperforms other tools on verification benchmarks~\cite{verification_as_chc}.

The runtimes of the three tools on the 56 instances of this parameterized benchmark are depicted in \cref{fig:escalating-graph}.
For all three tools, despite being based on very different automated analysis techniques, we observe that the respective runtimes increase significantly as the initial value of $max$ grows. Nevertheless, our approach still far outperforms both HyperQB and HyHorn for all tested parameters.
For large parameters, both HyperQB and HyHorn either get close to or exceed the 30 minutes timeout that we have set, whereas our tool finds the specification violation in less than 30 seconds.

\subsection{Summary}

The outcomes of this evaluation are promising.
First, the results obtained on all benchmarks show that our approach is able to find concrete counterexamples to relevant $\forall\exists$ hyperproperties, with convincing execution times: counterexamples are often found within fractions of a second.
Thus, we can positively answer questions Q1 and Q2: our approach is an effective and efficient way to detect $\forall\exists$ hyperproperty violations, fully automatically.

Regarding question Q3, our benchmarks contain examples that require to check whether a relations hold at multiple observation points, as well as examples where thousands of combinations of paths need to be checked. Importantly, for the more challenging benchmarks, other tools are either significantly slower, or they simply fail to detect violations without expert guidance. The fact that our prototype resists these stress tests with runtimes not exceeding a few seconds is encouraging, and calls for further development of this new approach.

\section{Related Work}\label{sec:rel-work}

\paragraph{Verification of hyperproperties.}
Dickerson et al.\ use a Hoare-style program logic to verify $\forall\exists$ hyperproperties~\cite{dickerson_rhle_2022}.
They automated their logic by generating verification conditions that are then discharged by an SMT solver. Upon verification failure, their implementation produces a counterexample.
However, as is common in all deductive verification approaches, loops in the input program present a major challenge.
In fact, the implementation of this program logic, ORHLE~\cite{dickerson_rhle_2022}, requires every loop to be annotated with relevant loop invariants.
In contrast, our approach based on symbolic execution can find counterexamples
even without loop invariants.

Beutner~\cite{beutner_2024} also proposed an automated approach to the verification of $\forall\exists$ hyperproperties based on a relational program logic. The associated tool, called ForEx, relies on a notion of \textit{parametrized strongest post-conditions} to automate the verification using symbolic execution techniques. Contrary to ORHLE and to our approach, ForEx cannot detect violations of hyperproperties.

Beutner and Finkbeiner developed a game-theoretic approach to the verification of $\forall\exists$-safety properties~\cite{beutner_software_2022} in infinite-state settings.
In their approach, the interaction between universally and existentially quantified traces is interpreted as a game between two agents.
If the existential player has a winning strategy, the property holds.
Such strategy-based verification methods are generally incomplete, i.e., verification might fail even though the respective property holds, and existing approaches to completeness do not scale beyond very small systems and constrained classes of properties~\cite{beutner_prophecy_2022,beutner_autohyper_2023}.
Additionally, when verification fails, game-based algorithms generally do not produce concrete counterexamples that witness the property violation.

Farzan et al. developed algorithms for the automated verification of $k$-safety properties in software based on the notion of \emph{reduction} \cite{farzan_automated_2019}.
Their approach simultaneously searches for a reduction (i.e., an alignment) of $k$ universally quantified programs, and an inductive proof compatible with this alignment.
The automatic search for appropriate reductions significantly improves the capability of the approach and widens the class of programs that can be automatically verified.

Yang et al.\ \cite{yang_lazy_2018} propose a lazy self-composition algorithm for information flow properties. The approach gradually refines an abstract (single copy) taint analysis by iteratively constructing a self-composed model. The approach is limited to secure information flow properties.

More recently, Itzhaky et al.~\cite{verification_as_chc} proposed to reduce the verification of hyperproperties, including those that require quantifier alternation, to Constrained Horn Clauses (CHC) satisfiability. They implemented their approach in a tool called HyHorn.
For some examples, HyHorn can refute hyperproperties fully automatically. However, as reported in our evaluation section, HyHorn has difficulties finding bugs in the more challenging examples.

\paragraph{Bounded model checking for hyperproperties.}
Hsu et al.\ apply bounded model checking to hyperproperties by reducing the verification task to checking the satisfiability of a quantified boolean formula (QBF)~\cite{groote_bounded_2021}.
In some sense, our algorithm lifts this approach to testing of infinite-state systems through the use of an SMT solver, yet we do not require the SMT solver's decision procedure to handle quantifier alternation in $\forall\exists$ hyperproperties. Instead, we eliminate the outermost universal quantifier(s) by explicitly enumerating traces using symbolic execution.

\paragraph{SMT solving and quantifier alternation.}
Naturally, our approach benefits from the significant advancements in the design and implementation of SMT solvers.
The SMT solver Yices introduced a dedicated algorithm for solving $\exists\forall$ queries in version 2.2 to aid in the synthesis of system parameters~\cite{dutertre_yices_2014}.
Their method, which we use for our experimental evaluation, provides termination guarantees for some underlying theories (including bitvectors and subsets of integer arithmetic).

\paragraph{Symbolic testing for hypersafety.}
Daniel et al.\ propose efficient methods to automatically test 2-safety security properties~\cite{lesly_binsec_2023,lesly_spectre_2021} such as constant-time, secret-erasure, and absence of Spectre attacks at the binary level.
Their approach is based on symbolic execution but tailored to these predefined properties, which allows them to develop important optimizations.
However, contrary to our approach, theirs cannot handle properties that require quantifier alternation.

\section{Conclusion}
\label{sec:conclusion}

We have presented a bounded semantics for the logic $\textrm{OHyperLTL}_\mathit{safe}$, and an algorithm capable of searching for counterexamples that witness the violation of $\forall\exists$-safety hyperproperties given user-defined input programs for the respective quantified trace variables.
We have demonstrated its effectiveness in locating specification violations by evaluating our implementation both against various examples from related work and against our own set of benchmarks, and we have shown experimentally that our approach outperforms existing tools in many cases. We have further characterized precisely under what circumstances our algorithm succeeds (as opposed to diverging).

It is worth noting that the counterexamples produced by the proposed algorithm consist of concrete traces for the universally quantified traces only.
While some SAT and SMT solvers can produce proofs of unsatisfiability, these automatically generated proofs are often difficult to relate to the high-level input semantics.
Future work includes an extension of the proposed algorithm that not only produces concrete traces for the universally quantified traces but also an explanation as to why these traces form a counterexample, likely in the form of a proof in terms of the high-level input semantics.

Our implementation supports imperative programs and utilizes the SMT theory of integer arithmetic.
We intend to extend its capabilities with support for other background theories in the future, which is possible because our method is theory-agnostic.

This novel approach to testing hyperproperties even in the presence of quantifier alternation lays the foundations for future research into fully automated hyperbug detection.
It is not only sound but even complete for a large class of programs.
Unlike many existing approaches, it is not limited to any particular hyperproperty, and our experimental evaluation has demonstrated its effectiveness on a wide range of specifications.
Given the promising results discussed in this paper, it seems likely that future enhancements will further improve the scalability of our approach, in which case it can be expected to become applicable to far more complex specifications and programs.

\section{Data Availability Statement}

The prototype implementation that is described in \cref{sec:eval} is available online:%
\begin{center}%
\href{https://github.com/tniessen/hyena0}{github.com/tniessen/hyena0}%
\end{center}%
The same repository also contains all input files that were used for the experimental evaluation.

\section*{Acknowledgements}

\begin{wrapfigure}{r}{3.5cm}%
\begin{center}%
\vspace{-1.4\baselineskip}
\noindent\euflag{!}{3.5\baselineskip}%
\end{center}%
\end{wrapfigure}

This work has received funding from the European Union's Horizon 2020 research and innovation programme under grant agreement No 101034440 and from the Vienna Science and Technology Fund (WWTF) [10.47379/VRG11005].
This work was also supported by the European Research Council (ERC) Grant HYPER (No. 101055412). 
Views and opinions expressed are however those of the authors only and do not necessarily reflect those of the European Union or the European Research Council Executive Agency. 
Neither the European Union nor the granting authority can be held responsible for them. 
A. Correnson carried out this work as a member of the Saarbrücken Graduate School of Computer Science.

\interlinepenalty=10000 
\bibliographystyle{ACM-Reference-Format}
\bibliography{references}

\end{document}

%% file: oopsla-auto-generated/orhle-benchmarks-table.tex
\begin{tabular}{ c l l c c c c | c }
  \toprule
  \textbf{Class} & \textbf{Type} & \textbf{Program} & \textbf{FO} & \textbf{Bug found} & \; \textbf{\# Combinations} \; & \;\textbf{Our tool} & \;\textbf{ORHLE}\\
  \midrule
  $\forall\exists$ & Refinement & \texttt{simple-nonrefinement} & Yes & \ok & 1 & 0.004 s & 0.221 s \\
  $\forall\exists$ & Other & \texttt{do-nothing} & Yes & \ok & 1 & 0.004 s & 0.226 s \\
  $\forall\forall\exists$ & GNI & \texttt{simple-leak} & Yes & \ok & 1 & 0.004 s & 0.225 s \\
  $\forall\exists$ & Other & \texttt{draw-once} & Yes & \ok & 1 & 0.004 s & 0.223 s \\
  $\forall\forall\exists$ & GNI & \texttt{nondet-leak2} & Yes & \ok & 2 & 0.004 s & 0.235 s \\
  $\forall\forall\exists$ & GNI & \texttt{smith1} & Yes & \ok & 2 & 0.009 s & 0.221 s \\
  $\forall\forall\exists$ & GNI & \texttt{nondet-leak} & Yes & \ok & 2 & 0.009 s & 0.235 s \\
  $\forall\forall\exists$ & Delimited Release & \texttt{parity-no-dr} & Yes & \ok & 2 & 0.009 s & 0.226 s \\
  $\forall\forall\exists$ & Delimited Release & \texttt{wallet-no-dr} & Yes & \ok & 2 & 0.009 s & 0.233 s \\
  $\forall\exists$ & Refinement & \texttt{conditional-nonrefinement} & Yes & \ok & 4 & 0.015 s & 0.230 s \\
  $\forall\exists$ & Refinement & \texttt{add3-shuffled} & Yes & \ok & 6 & 0.024 s & 0.268 s \\
  $\forall\forall\forall\exists$ & Delimited Release & \texttt{conditional-no-dr} & Yes & \ok & 8 & 0.035 s & 0.232 s \\
  $\forall\forall\exists$ & Delimited Release & \texttt{median-no-dr} & Yes & \ok & 4 & 0.042 s & 0.517 s \\
  $\forall\forall\forall\exists$ & Delimited Release & \texttt{conditional-leak} & Yes & \ok & 48 & 0.187 s & 0.237 s \\
  $\forall\exists$ & Refinement & \texttt{loop-nonrefinement} & No & \ko & N/A & $\infty$ & 0.258 s \\
  \bottomrule
\end{tabular}

%% file: oopsla-auto-generated/our-benchmarks-table.tex
\begin{tabular}{ c l c c c c }
  \toprule
  \textbf{Class} & \textbf{Program} & \textbf{Bug found} & \textbf{\# Observations} & \textbf{\# Combinations} & \textbf{Runtime} \\
  \midrule
  $\forall\exists$ & \texttt{even\_odd} & \ok & 1 & 1 & 0.002 s \\
  $\forall\exists$ & \texttt{factor2} & \ok & 2 & 2 & 0.004 s \\
  $\forall\exists$ & \texttt{for\_loop\_simple} & \ok & 1 & 2 & 0.026 s \\
  $\forall\exists$ & \texttt{linear\_equation} & \ok & 22 & 22 & 0.061 s \\
  $\forall\exists$ & \texttt{monotonic\_increase} & \ok & 7 & 7 & 0.071 s \\
  $\forall\exists$ & \texttt{escalating\_inl} & \ok & 7 & 40 & 0.230 s \\
  $\forall\exists$ & \texttt{escalating\_2} & \ok & 7 & 747 & 0.259 s \\
  $\forall\forall\exists$ & \texttt{secret\_pin\_leak} & \ok & 8 & 11 & 0.265 s \\
  $\forall\exists$ & \texttt{escalating} & \ok & 7 & 1195 & 0.350 s \\
  $\forall\exists$ & \texttt{escalating\_3} & \ok & 7 & 1707 & 0.445 s \\
  $\forall\forall$ & \texttt{obs\_determinism} & \ok & 4 & 86 & 1.081 s \\
  $\forall\exists$ & \texttt{no\_primes\_above\_31397} & \ok & 1 & 201 & 2.481 s \\
  $\forall\forall\exists$ & \texttt{secret\_pin\_leak\_2} & \ok & 3 & 248 & 5.303 s \\
  $\forall\exists$ & \texttt{exponential\_branching\_1} & \ok & 1 & 1024 & 5.566 s \\
  $\forall\exists$ & \texttt{exponential\_branching\_2} & \ok & 1 & 2048 & 10.933 s \\
  \bottomrule
\end{tabular}